\documentclass[useAMS,usenatbib]{mn2e}

\usepackage[latin1]{inputenc}
\usepackage[english]{babel}
\usepackage{epsfig}
\usepackage{amssymb}
\usepackage{times}
\usepackage{amsmath}
\usepackage{graphicx}
\usepackage{epstopdf}
\usepackage{mathrsfs}
\usepackage{tabularx}
\usepackage{rotating}
\usepackage{longtable,lscape}
\usepackage{natbib}
\usepackage{multirow}

\newcommand\ion[2]{#1$\;${\scshape{#2}}}%

\def\apjs{ApJS}

\def\aap {A\&A}

\title[A multi-wavelength continuum characterization of high-z BAL QSOs]{A multi-wavelength continuum characterization of high-redshift broad absorption line quasars}

\author[D. Tuccillo et al.]{D. Tuccillo,$^{1, 2}$\thanks{E-mail:diego.tuccillo@obspm.fr} 
G. Bruni,$^{3}$
M. A. DiPompeo,$^{4,5}$
M. S. Brotherton,$^{5}$
A. Pasetto,$^{3}$ \newauthor
A. Kraus,$^{3}$
J. I. Gonz\'alez-Serrano,$^{2}$
and K.-H. Mack$^{6}$
\\
$^{1}$GEPI, Observatoire de , CNRS, Universit\'e  Diderot, 61, Avenue de l'Observatoire F-75014, , France\\
$^{2}$Instituto de F\'isica de Cantabria (CSIC-Universidad de Cantabria), Avda. de los Castros s/n, E-39005 Santander, Spain\\
$^{3}$Max Planck Institute for Radio Astronomy, Auf dem Huegel, 69, D-53121  Bonn, Germany\\
$^{4}$Department of Physics and Astronomy, Dartmouth College, 6127 Wilder Laboratory, Hanover, NH 03755, USA \\
$^{5}$Dep. of Physics and Astronomy 3905, University of Wyoming, 1000 East University, Laramie, WY 82071, USA\\
$^{6}$INAF-Istituto di Radioastronomia, via Piero Gobetti, 101, I-40129 Bologna, Italy\\
}

\begin{document}

\date{Accepted for publication on MNRAS, 2017 February 6}

\pagerange{\pageref{firstpage}--\pageref{lastpage}} \pubyear{2011}   

\maketitle

\label{firstpage}

\begin{abstract}
We present  the results of a multi-wavelength study of a sample of high-redshift Radio Loud (RL) Broad Absorption Line (BAL) quasars. This way we extend to higher redshift previous studies on the radio properties, and broadband optical colors of these objects. We have selected a sample of 22 RL BAL quasars with $3.6 \le z \le 4.8$ cross-correlating the FIRST radio survey with the SDSS. Flux densities between 1.25 and 9.5 GHz have been collected with the JVLA and Effelsberg-100m telescopes for 15 BAL and 14 non-BAL quasars used as comparison sample. 
We determine the synchrotron peak frequency, constraining their age. A large number of GigaHertz Peaked Spectrum (GPS) and High Frequency Peakers (HFP) sources has been found in both samples (80\% for BAL and 71\% for non-BAL QSOs), not suggesting a younger age for BAL quasars. The spectral index distribution provides information about the orientation of these sources, and we find statistically similar distributions for the BAL and non-BAL quasars in contrast to work done on lower redshift samples.  Our sample may be too small to convincingly find the same effect, or might represent a real evolutionary effect based on the large fraction of young sources.
We also study the properties of broadband colors in both optical (SDSS) and near- and mid-infrared (UKIDSS and WISE) bands, finding that also at high redshift BAL quasars tend to be optically redder than non-BAL quasars. However, these differences are no more evident at longer wavelength, when comparing colors of the two samples by mean of the WISE survey.
\end{abstract}



\section{Introduction}
Intrinsic absorption lines are a common feature in quasar (QSO) spectra, and they are likely produced by outflowing winds along the line of sight that are launched from the accretion disk around the central supermassive black hole (\citealt{Murray_1995}; \citealt{Proga_2000}). Broad absorption lines (BALs) in ultraviolet and visible spectra of quasars are the most spectacular manifestation of such outflows, with absorption velocity width larger than 2000 $\rm{km/s}$ (\citealt{Weymann}). BAL quasars are often classified into three subtypes depending on the presence of absorption lines in specified transitions: (1) High-ionization (Hi) BAL quasars show absorption lines in high-ionization transitions such as \ion{Si}{iv}, and \ion{C}{iv}; (2) Low-ionization (Lo) BAL quasars possess \ion{Mg}{ii}  and/or \ion{AI}{iii}  absorption lines, in addition to the high-ionization transitions. (3) Iron low-ionization (FeLoBAL) BAL quasars show additional absorption from excited states of \ion{Fe}{ii}  and  \ion{Fe}{iii} (e.g., \citealt{Hall_2002}). About 15\% (\citealt{Knigge_2008}) of the whole population of quasars are classified as BAL, although some authors \citep{Ganguly_2007} suggest that the intrinsic fraction could be much higher.  

Powerful outflows can enrich the quasar host galaxy interstellar medium, contributing to a variety of feedback effects such as regulating host galaxy star formation rates (\citealt{Hopkins_2010}), and limiting quasar lifetimes by removing fuel from the nuclear regions (\citealt{Silk_1998}). Understanding the physics behind BAL outflows may be fundamental to disclose the connection between the evolution of quasars and their host galaxies. 
According to the \textit{orientation scenario}, BAL QSOs are normal quasars seen at high inclination angles, and the outflows at the origin of the BAL features are located near the equatorial plane of the accretion disk. This model fits naturally in the AGN unification models (\citealt{Antonucci_1993}; \citealt{Urry_1995}) and it is supported by the claim that both BAL and non-BAL QSOs appear to be drawn by the same parent population \citep{Reichard_2003}. 
However, the orientation scenario is contradicted from the studies on the radio-loud (RL) BAL quasars, studied for the first time in \cite{Brotherton_1998} and in \cite{Becker1}. In fact an edge-on geometry should lead to lobe-dominated emission and thus steep radio spectrum but, using large samples of RL BAL QSOs (\citealt{Montenegro}, \citealt{Fine_2011}, \citealt{DiPompeo_2011}, \citealt{Bruni_2012}), it has been confidently established that these objects have a variety of spectral indices for their radio Spectral Energy Distribution (SED hereafter). On the light of these results, a pure orientation scenario is clearly inadequate to explain the full extent of the BAL phenomena, especially considering that most of the continuum and emission line properties of RL and RQ BALs appear statistically identical and therefore suggest that the results obtained on RL QSOs quasars can be extended to their RQ counterparts \citep{Rochais_2014}. The other main model proposed to explain the nature of BAL QSOs is \textit{evolutionary model} as discussed for example by \cite{Lipari_2006}. According to this, BAL QSOs represent a particular evolutionary stage of quasars, during which absorbing material with a high covering fraction is being expelled from the central regions of the quasar. The radio-loud systems may be associated with the later stages of evolution \citep{Becker0}, when jets have removed the clouds responsible for the generation of BALs. 
The early claim that BAL QSOs have redder continua has been confirmed by a number of different studies in the UV/optical domain (\citealt{Weymann}; \citealt{Sprayberry_1992}; \citealt{Brotherton_2001}; \citealt{Reichard_2003}; \citealt{Urrutia_2009}) and it has been pointed out in support of the evolutionary scenario because it is in agreement with the dustier environment expected in a young AGN. However, the results of the studies at higher wavelength are controversial (\citealt{Gallagher_2007}, \citealt{DiPompeo_2013}) 
It has also been proved that RL BAL QSOs show a variety of possible radio morphologies (\citealt{Montenegro}; \citealt{DiPompeo_2011}; \citealt{Bruni_2012}, \citealt{Bruni_2013}) and are not always unresolved, as it may be expected for young radio sources \citep{Fanti_1990}.

One of the most successful and studied BAL catalogs is the one from \cite{Gibson_2009} that includes 5,039 BAL QSOs, and the most recent Quasar Catalog (Paris et al, in prep) based on the Sloan Digital Sky Survey Data Release 12 (SDSS DR12) includes $\sim$30,000 BAL QSOs. In spite of these large samples of BAL quasars and the proved interest of the radio-studies on this population, there is still a lack of observational information about RL BAL QSOs at high redshifts. In fact these studies are hampered by the lower radio flux density collected at higher redshift and strongly bounded by the scarce number of RL BAL QSOs known at $z > 3$.  RL QSOs account only for 8-13\% of  QSOs  (\citealt{Ivezic_2002};  \citealt{Jiang_2007};  \citealt{Balokovic_2012}), and the density of quasars is a strong function of redshift that peaks at $z \sim 2-3$ and declines exponentially toward lower and higher redshift (e.g., \citealt{Boyle_2000}).  Finally the identification of BAL QSO is limited in SDSS at $z \lesssim 4.9$, since absorption in high-ionization species like \ion{C}{iv} $\, 1550 \, \AA$ is required for the classification of a quasar as a BAL. As a result the limit in redshift for the work of \cite{Montenegro} was $z=3.4$ while the samples of \cite{DiPompeo_2011} and \cite{Bruni_2012} had a maximum of $z=3.42$ and $3.38$, respectively. Extension of the samples to higher redshift is an important test to look for evolutionary trends in BAL properties. Indeed, \citet{Allen_2011} already found a dependence on redshift of the BAL QSO fraction, the latter decreasing by a factor 3.5$\pm$0.4 from redshifts z$\sim$4.0 down to z$\sim$2.0, and they interpret this as an evolutionary behavior, not reproducible by the orientation model alone. 

%

In this work we aim to complete the radio properties studies carried out by previous authors, filling the gap of information $ 3.6 \le z \le 4.8$. For this purpose, we selected and collected the radio flux densities of a sample of 15 RL BAL QSOs and 14 normal RL QSOs matched in redshift and magnitude. We observed these QSOs at 6 different frequencies in order to reconstruct their SED.  The age and the orientation of the radio jet is estimated from the SED. In particular the spectral index is used as a statistical indicator of the orientation of the jet with respect to the observer's line of sight, thus helping in verifying a possible preferred orientation. The number of GigaHertz Peaked Spectrum (GPS) sources is used as indicative of whether these objects are typically younger than a sample of "normal" QSOs. Imaging synthesis, possible with the interferometric observations, is used to give indications about the extension and morphology of these sources.

We also used a larger sample of 22 RL high-z BAL QSOs to compare their broadband optical colors with those of 113 RL QSOs with  $ 3.6 \le  z \le 4.8$. This way we investigate the question of the continuum differences between BALQSOs and non-BALQSOs, comparing our results with those from other authors (\citealt{Menou_2001}, \citealt{Tolea_2002}, \citealt{Reichard_2003}) at lower redshift or for samples of radio-quiet (RQ) BALs.

In the same range of redshift  $ 3.6 \le  z \le 4.8$ we also used a larger sample of 22 RL high-z BAL QSOs to compare their broadband optical colors with those of 113 normal RL QSOs with comparable redshift. This way we investigate the question of the continuum differences between BALQSOs and non-BALQSOs, comparing our results with those from other authors (\citealt{Menou_2001}, \citealt{Tolea_2002}, \citealt{Reichard_2003}) at lower redshift or for samples of radio-quiet (RQ) BALs.

 In this paper we  adopt the $\Lambda$CDM cosmology with $\Omega_{\Lambda} =0.7$, $\Omega_m =0.3$, and $H_0 =70 \, km \, s^{-1} \, Mpc^{-1}$.

\begin{table*}
\centering
\caption{Sample of 22 RL BAL selected from sources matched in FIRST and SDSS DR7}
\renewcommand\tabcolsep{4pt}
\begin{tabular}{c c c c c c c c c r c }
\hline
\multicolumn{1}{c}{SDSS ID} &
\multicolumn{1}{c}{ID} &
\multicolumn{1}{c}{RA} &
\multicolumn{1}{c}{DEC} &
\multicolumn{1}{c}{$i_{AB}$} &
\multicolumn{1}{c}{$S_{1.4\rm{GHz}}$} &
\multicolumn{1}{c}{z} &
\multicolumn{1}{c}{log$_{10}P_{1.4\rm{GHz}}$} &
\multicolumn{1}{c}{log$_{10}R$} &
\multicolumn{1}{c}{AI*}  &
\multicolumn{1}{c}{Note} \\
\multicolumn{1}{c}{} &
\multicolumn{1}{c}{} &
\multicolumn{2}{c}{(J2000)} &
\multicolumn{1}{c}{} &
\multicolumn{1}{c}{(mJy)} &
\multicolumn{1}{c}{} &
\multicolumn{1}{c}{(W/Hz)} &
\multicolumn{1}{c}{(mag)} &
\multicolumn{1}{c}{(km/s)} &
\multicolumn{1}{c}{}  \\
\multicolumn{1}{c}{(1)} &
\multicolumn{1}{c}{(2)} &
\multicolumn{2}{c}{(3)} &
\multicolumn{1}{c}{(4)} &
\multicolumn{1}{c}{(5)} &
\multicolumn{1}{c}{(6)} &
\multicolumn{1}{c}{(7)}  &
\multicolumn{1}{c}{(8)}  &
\multicolumn{1}{c}{(9)}  &
\multicolumn{1}{c}{(10)} 
\\
\hline
J000051.57+001202.5    & 0000+00		&  00:00:51.57		& +00:12:02.5	& 19.95		&	2.99      & 4.00	&   26.15	& 2.16	&	3762.29		& (a), (g), (p)	\\
J074738.49+133747.3    & 0747+13  	&  07:47:38.49		& +13:37:47.3	& 19.15 		&	6.62      & 4.04	&   26.53	& 2.13	& 	5879.09    	&			\\
J094003.03+511602.7    &     -               	&  09:40:03.03   		& +51:16:02.7  	& 18.77 		&	13.91    &  3.60	&   26.76	& 2.33	& 	166.42  		&   			\\
J100645.59+462717.2    & 1006+46     	&  10:06:45.59   		& +46:27:17.2   	& 19.81 		&	6.32      & 4.44	&   26.58 	& 2.65 	& 	185.31 		& 			\\
J102343.13+553132.4    & 1023+55     	&  10:23:43.13 		& +55:31:32.4   	& 19.31 		&	2.80      & 4.45	&   26.23 	& 2.25 	& 	4820.07 	&(a), (g) 		 \\
J103601.03+500831.8    & 1036+50     	&  10:36:01.03 		& +50:08:31.8	& 19.20 		&    9.22      & 4.47	&   26.75 	& 2.65	& 	1737.88 	& (a), (g) 	\\
J110946.44+190257.6    & 1109+19		&  11:09:46.44 		& +19:02:57.6     & 20.03		&    6.95      & 3.67    & 26.47		& 2.46	& 	1678.28 	& (p)		 	\\
J111055.22+430510.1    & 1110+43    	&  11:10:55.22 		& +43:05:10.1     & 18.21 	&	1.21      & 3.82    & 25.75		& 1.11 	& 	193.72 		&  	(p)	 	\\
J112938.73+131232.3    & 1129+13     	&  11:29:38.73 		& +13:12:32.3     & 18.76 	& 	1.33      & 3.61    & 25.74		& 1.22  	& 	206.46  		& (p)		 	\\
J113330.91+380638.2    & 	 	-		&  11:33:30.91		& +38:06:38.2 	 & 19.66 	&     0.87 	& 3.63	 & 25.56		& 1.42    &	176.29 		& (g), (p)		 \\
J115731.67+225726.4    & 1157+22     	&  11:57:31.67 		& +22:57:26.4     & 20.14 	& 	3.81      & 3.92    & 26.26		& 2.32  	& 	979.66 		&  (p)		 	\\
J120447.15+330938.7    &        -           	&  12:04:47.15 		& +33:09:38.7     & 18.38 	& 	0.92      &  3.61   & 25.58		& 1.25   	& 	6534.97 	& (a), (g), (p) 	\\
J124658.83+120854.7    & 	-		&  12:46:58.83		& +12:08:54.72	 & 19.86 	& 	1.44 	&  3.80 	 & 25.82 	&   1.76	& 	1519.70 	& (a), (g), (p) 	\\
J130348.94+002010.5    & 	-		&  13:03:48.94 		& +00:20:10.51	 & 18.66 	& 	0.99 	&  3.65	 & 25.62 	&   1.15  & 	1032.38 	& (a), (g), (p)	 \\
J133234.18+000921.7    & 1332+00     	&  13:32:34.18 		& +00:09:21.7     & 20.30 	& 	1.49      & 3.66    & 25.80		& 1.95	& 	149.60 		& (t) 	  	\\
J134428.55+625608.2    & 1344+62     	&  13:44:28.55 		& +62:56:08.2     & 19.23 	& 	2.58      & 3.67    & 26.04 	& 1.77 	& 	4967.82 	& (t), (g)  	\\
J134854.37+171149.6    & 1348+17     	&  13:48:54.37 		& +17:11:49.6     & 18.91 	& 	1.89      & 3.62    & 25.90 	& 1.52  	& 	356.61 		&  (p)		 	\\
J135554.56+450421.0    &  	-		&  13:55:54.56 		& +45:04:21.0	& 19.31 		& 	2.07  	& 4.09	 & 26.03  	& 1.62   & 	330.64 		& (g) 		\\
J150643.81+533134.5    & 1506+53     	&  15:06:43.81 		& +53:31:34.5     & 18.77 	& 	14.63    & 3.79    & 26.82 	& 2.34   	& 	166.89 		& (t) 		\\
J151146.99+252424.3    & 1511+25     	&  15:11:46.99 		& +25:24:24.3     & 19.76 	& 	1.39      & 3.72    & 25.78 	& 1.72  	& 	3034.72 	& (a), (p) 		\\
J161716.49+250208.1    &  	-		&  16:17:16.49 		& +25:02:08.1	 & 19.83 	& 	2.35	&  3.94	 & 26.06		& 1.95   & 	4439.58 	& (a), (g), (p)	 \\
J165913.23+210115.8    & 1659+21     	&  16:59:13.23 		& +21:01:15.8     & 20.11 	& 	28.81    & 4.78    & 27.30 	& 3.37 	&	787.91 		& 		 	\\
\hline
\label{BALTable}
\end{tabular}

\medskip

\begin{flushleft}
\small  The columns give the following: (1) SDSS object-ID; (2) shortened name assigned to source for the radio observations, sources without this ID were not used for our radio studies ; (3) SDSS J2000 coordinates; (4) SDSS $i$-magnitude corrected for galactic extinction; (5) FIRST integrated radio flux density; (6)QSO redshift; (7) radio luminosity at $1.4$ GHz; (8) R-parameter of radio-loudness (Kellermann et al. 1989);  (9) Modified absorption index AI* (Bruni et al. 2012);  (10) note indicating if the QSO is classified as BAL in previous catalogs, (a) Allen et al. 2011, (g) Gibson et al. 2009 (t) Trump et al. 2006 (p) P\^aris et al. 2012.
\end{flushleft}
\end{table*}


\begin{table*}
\centering
\caption{Comparison sample of non-BAL QSOs observed with the Effelsberg-100m single dish and/or the JVLA interferometer.}
\label{Table_comparison}
\renewcommand\tabcolsep{4pt}
\begin{tabular}{c c c c c c c c c}
\hline
\multicolumn{1}{c}{SDSS ID} &
\multicolumn{1}{c}{ID} &
\multicolumn{1}{c}{RA} &
\multicolumn{1}{c}{DEC} &
\multicolumn{1}{c}{$i_{AB}$} &
\multicolumn{1}{c}{$S_{1.4\rm{GHz}}$} &
\multicolumn{1}{c}{z} &
\multicolumn{1}{c}{log$_{10}P_{1.4\rm{GHz}}$} &
\multicolumn{1}{c}{log$_{10}R$} 
\\
\multicolumn{1}{c}{} &
\multicolumn{1}{c}{} &
\multicolumn{2}{c}{(J2000)} &
\multicolumn{1}{c}{} &
\multicolumn{1}{c}{(mJy)} &
\multicolumn{1}{c}{} &
\multicolumn{1}{c}{(W/Hz)} &
\multicolumn{1}{c}{(mag)} 
\\
\multicolumn{1}{c}{(1)} &
\multicolumn{1}{c}{(2)} &
\multicolumn{2}{c}{(3)} &
\multicolumn{1}{c}{(4)} &
\multicolumn{1}{c}{(5)} &
\multicolumn{1}{c}{(6)} &
\multicolumn{1}{c}{(7)}  &
\multicolumn{1}{c}{(8)}  
\\
\hline
J030025.23+003224.2    &   0300+00   &   03:00:25.23   & +00:32:24.2    & 19.68  & 7.69       & 4.18    & 26.62 & 2.31   \\
J083322.50+095941.2    &   0833+09   &   08:33:22.50   & +09:59:41.2    & 18.60  & 125.76   & 3.73    & 27.74 & 3.18    \\
J084044.19+341101.6    &   0840+34	&   08:40:44.19   & +34:11:01.6    & 19.58  & 13.59     & 3.89    & 26.81 & 2.64   \\
J090129.23+104240.4    &   0912+10	&   09:01:29.23   & +10:42:40.4    & 20.08  & 2.14       & 3.96    & 26.02 & 2.01    \\
J091824.38+063653.4    &   0918+06   &   09:18:24.38   & +06:36:53.4    & 19.18  & 26.50     & 4.19    & 27.16 & 2.88        \\
J101747.76+342737.9    &   1017+34   &   10:17:47.76   & +34:27:37.9    & 19.99  & 2.64       & 3.69    & 26.06 & 2.02       \\
J110201.91+533912.6    &   1102+53   &   11:02:01.91   & +53:39:12.6    & 20.31  & 5.57       & 4.30    & 26.50 & 2.51       \\
J112530.50+575722.7    &   1125+57   &   11:25:30.50   & +57:57:22.7    & 19.44  & 2.99       & 3.68    & 26.11 & 1.85       \\
J115045.61+424001.1    &   1150+42   &   11:50:45.61   & +42:40:01.1    & 19.87  & 1.51       & 3.87    & 25.85 & 1.72       \\
J124943.67+152707.1    &   1249+15   &   12:49:43.67   & +15:27:07.1    & 19.05  & 2.00       & 3.99    & 26.00 & 1.61      \\
J131121.32+222738.7    &   1311+22   &   13:11:21.32   & +22:27:38.7    & 20.19  & 6.53       & 4.61    & 26.62 & 2.84       \\
J142326.48+391226.3    &   1423+39   &   14:23:26.48   & +39:12:26.3    & 20.04  & 6.51       & 3.92    & 26.50 & 2.46     \\
J144643.37+602714.4    &   1446+60   &   14:46:43.37   & +60:27:14.4    & 19.74  & 1.80       & 3.78    & 25.91 & 1.77      \\
J161105.65+084435.4    &   1611+08   &   16:11:05.65   & +08:44:35.4    & 18.84  & 8.82       & 4.54    & 26.74 & 2.31       \\
\hline
\end{tabular}

\medskip
\begin{flushleft}
\small  The columns give the following: (1) SDSS object-ID; (2) shortened name assigned to source for the radio observations ; (3) SDSS J2000 coordinates; (4) SDSS $i$-magnitude corrected for galactic extinction; (5) FIRST integrated radio flux density; (6) QSO redshift; (7) radio luminosity at $1.4$ GHz; (8) R-parameter of radio-loudness (Kellermann et al. 1989)
\end{flushleft}
\end{table*}

\section{The high-z RL BAL quasar sample}
We selected a sample of spectroscopically identified RL BAL QSOs with $ 3.6 \le z \le 4.8$ and detected in both SDSS DR7 and FIRST. The range of redshifts was chosen in order to mostly overlap the one used in \cite{Tuccillo}, where we used a Neural Networks machine method to select RL quasars candidates in the redshift range $ 3.6 \le z \le 4.4$. This selection method was proven to be highly complete (97\%) and it allowed the spectroscopic identification of 15 QSOs missed in SDSS DR7 (\citealt{Carballo_2008}; \citealt{Tuccillo}). The reduced biases in the selection lead to a very accurate estimation of the optical luminosity function (LF) of radio-loud quasar in this range of redshift. Exploring a similar range of redshift,  we search for BAL QSOs within a very complete spectroscopic sample of RL QSOs. Moreover, the careful determination of the quasar LF, assures that the RQ and the RL QSOs populations do not evolve differently in this range of redshift. Therefore any possible difference found in this research between BAL and non BAL QSOs, are not a-priori biased by different evolutions between RL and RQ populations. 

A simple one-to-one match between FIRST and SDSS will miss double-lobe QSOs without detected radio cores. \cite{deVries_2006} found that for a sample of 5515 FIRST- SDSS QSOs with radio morphological information within 450 arcsec, the fraction of FIRST-SDSS double-lobe QSOs with undetected cores is 3.7 per cent. Since the starting samples of SDSS QSOs in  \cite{deVries_2006} and in this work obey similar SDSS selection criteria, we estimate that our sample is similarly affected for this source of incompleteness.

All the sources of our sample are radio-loud in agreement with both the main definitions adopted in literature to consider a quasar as "radio loud". In particular they meet the criteria  based on the total radio luminosity of the source adopted by \cite{Gregg_1996}, i.e. log$ \, P_{1.4, \rm GHz} \rm{(W/Hz)} >25.5$. They also have $R>10$, commonly used threshold to define radio-loudness on the basis of the $R$-parameter, defined as the rest-frame ratio of the  monochromatic 6-cm (5 GHz) and 2500\AA \, flux densities (\citealt{Kellermann_1989} ; \citealt{Stocke_1992}). We also note that, given the high-z of our sample of BAL QSOs, we can not measure the transition lines needed to distinguish our BAL QSOs sample in Hi-,Lo- or FeLo- BAL QSOs, and therefore our sample is likely to include more than just one subclass of BALs. 

In section \ref{BALselection_sec} we give the details of our selection of BAL QSOs.

\subsection{BAL QSOs selection criteria}
\label{BALselection_sec}

We started our selection considering the sample of 222,517 sources obtained cross matching each source of the FIRST survey (2003 April 11 version), not flagged as possible sidelobe or nearby bright source ($\sim 3.6\%$ of the sources in the FIRST catalogue), with the closest optical object in the "PhotoPrimary" view of the SDSS DR7 catalogue within a $1\farcs5$ radius. In this sample there is no selection by radio flux density or radio morphology other than the requirement that the radio source has at least a weak core component. From this sample we discarded all the sources not classified as "point-like" or tagged with the "fatal" error flags by the SDSS pipelines. In this way we pre-selected a sample of 36,267 sources. 

At this point we searched for all the available optical spectra for the sources of this sample, either from the SpecObj view of SDSS-DR7 or from the 5th edition of the SDSS Quasar Catalogue (DR7 QSO Catalogue; \citealt{Schneider}). We integrated the sample with all the new quasars discovered with our neural-network selection strategies (\citealt{Tuccillo}).  We  visually inspected all the spectra of the sources classified as QSOs at $z> 3$ in the DR7 QSO catalogue or in SDSS-DR7, looking for broad absorption trough in the \ion{Ci}{iv} line. 

Finally we measured the absorption in the \ion{Ci}{iv} for a rigorous and homogeneous selection of the final sample of BAL QSOs. In literature there is more than one metric used to separate BALQSOs and non-BAL quasars on the basis of the measured absorption. The most widely used definitions are: (i) the \textit{balnicity index} (BI, first presented by \citealt{Weymann}); (ii) the \textit{absorption index} (AI) (\citealt{Hall_2002}; \citealt{Trump}), designed to include narrower troughs than the BI; (iii) and the \textit{modified balnicity index} $BI_0$ (\citealt{Gibson_2009}), which extends the integration region to zero velocity. In this work, to discriminate between BAL and not-BAL QSOs, we choose to apply the same criteria used in \cite{Bruni_2012},  in order to  compare our results with their studies at lower redshift. Therefore we calculated the absorption index (AI), as defined  by \cite{Hall_2002}:
\begin{equation}
AI = \int_{0 \, \rm{km/s}}^{25000 \, \rm{km/s}} (1-\frac{f(\nu)}{0.9} ) C d\nu
\end{equation}
\noindent  where $f(\nu)$ is the continuum-normalized flux (unsmoothed whenever possible) as a function of velocity $\nu$, relative to the line centre. We integrate the spectral region between the peaks of the \ion{Ci}{iv} and  \ion{Si}{iv} emission lines to up to $25 000 \, \rm{km/s}$ from the former.  The constant $C$ is posed to be equal to one when $f(\nu)$ is less than 0.9 for at least $1000 \, \rm{km/s}$ (as in \citealt{Trump}), and it is posed to be equal to zero elsewhere.  We considered as genuine BAL QSOs only objects with an $AI>100 \, \rm{km/s}$. 
Applying this criteria we ended-up with a sample of 22 BAL QSOs within $ 3.6 \le z \le 4.8$. Comparing (see Table  \ref{BALTable})  this sample with larger catalogues of BALs that are also based on the SDSS and therefore overlapping the sky area of our sample, we verified that  9 of the sources included in our sample are included also in the catalogue of \cite{Allen_2011}, 10 are included also in the Gibson catalogue (\citealt{Gibson_2009}),  and 3 in the catalogue from \cite{Trump}. 12  QSOs are also classified as BAL in the DR12 QSO Catalog \cite{Paris_2012}, published after our selection.  9 BALs of our sample were not classified as such in literature at the moment of the selection, and 4 have been identified as BAL QSOs in this work for the first time.

\begin{figure} 
\centering 
\includegraphics[width=80mm]{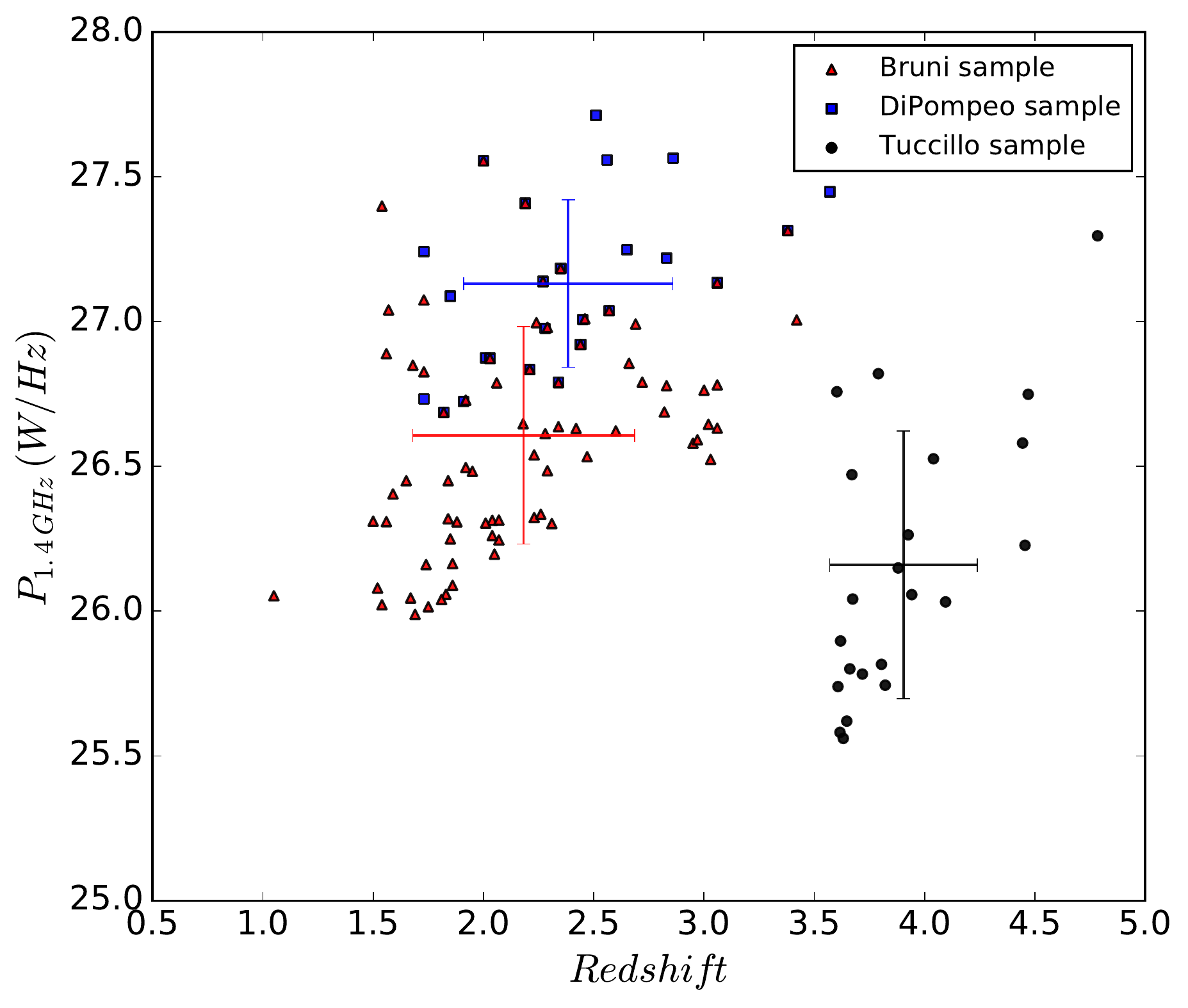}%
\qquad
\qquad 
\includegraphics[width=80mm]{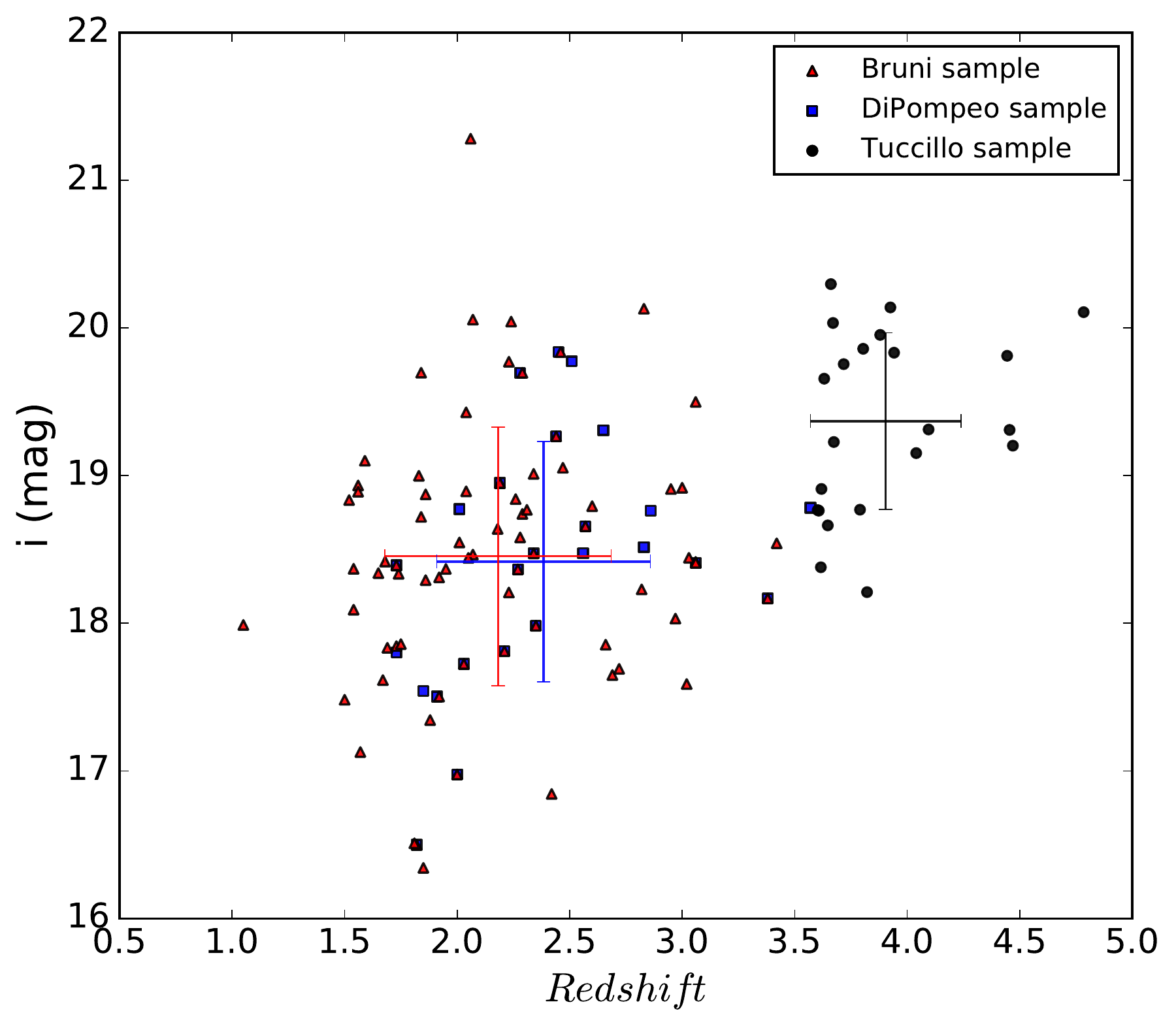} 
\qquad
\qquad 
\includegraphics[width=80mm]{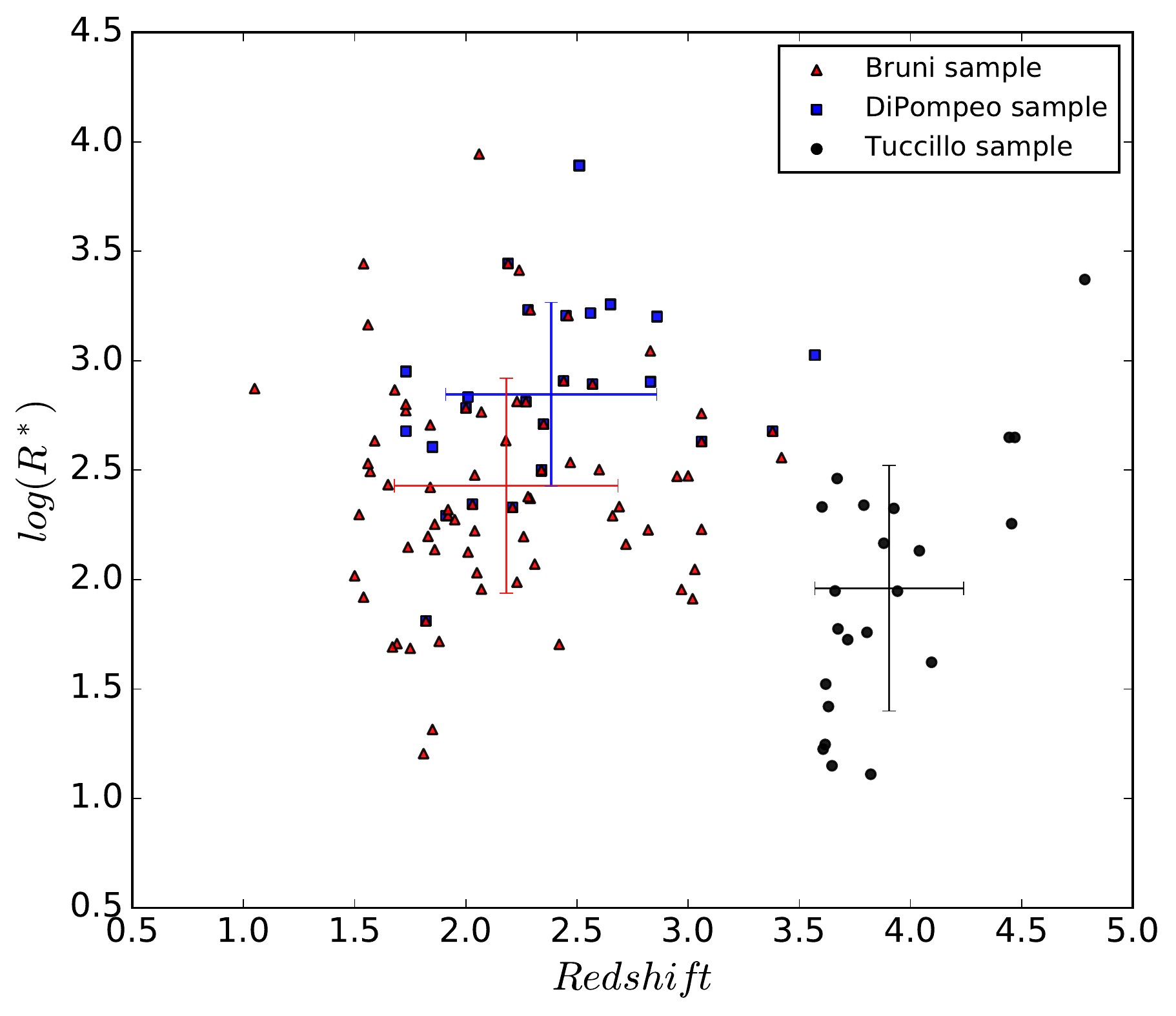} 
\caption{\small  We show, as a function of the redshift, the distributions in radio luminosity (at  $\log_{10} P_{1.4\rm{GHz}} \, (\rm{W/Hz})$ ), i-magnitude, and radio-loudness $R$ for the 25 RL BAL QSOs (red triangles) used in Bruni et al. (2012), for the 74  RL BAL QSOs (blue squares) used in DiPompeo et al. (2011) and for the 22 RL BAL QSOs (black dots) used in this work. The large crosses with 1-$\sigma$ error bar, represent the location of the mean for each of the plotted samples.  The sample studied in this work was selected in order to extend the previous cited studies at higher redshift and at lower optical and radio fluxes,  probing new parameter space. See section \ref{BALselection_sec}}
\label{BAL_Br_DiP_Tu}
\end{figure}

Table \ref{BALTable} provides a catalog of the 22 BAL QSOs of our sample. In the table we include the measure of the AI for each BAL QSOs and we indicate which of these sources were used for the radio studies that we will discuss in the next section. As a whole, and as shown in Fig. \ref{BAL_Br_DiP_Tu}, our sample of 22 RL BAL quasars  extend the samples used in \cite{DiPompeo_2011} and \cite{Bruni_2012}, for the redshift range covered (out to $z=4.8$) and extend to lower optical and radio flux densities.

\section{Radio observations and data reduction}
With the aim of repeating at high-z radio-studies similar to those of \cite{Montenegro}, \cite{DiPompeo_2011} and \cite{Bruni_2012} we originally proposed to observe the 22 BAL QSOs of our sample and a "comparison sample" of 22 unabsorbed QSOs in multiple frequency bands, including 1.4 GHz (L-band), 4.9 GHz (C-band), 8.4 GHz (X-band), and 22 GHz (K-band) at the 100-m Effelsberg Telescope and at the EVLA radio-array in the A configuration. Not all sources of the samples were observed, either because they were too faint in radio or for technical problems related to bad weather. Nonetheless the final observed sample, composed of 15 (out of the 22) BAL QSOs, and 14 QSOs of the "comparison sample", provided sufficient statistic to complete the planned studies.

In Table \ref{BALTable} of section \ref{BALselection_sec} we have indicated the 15 BAL QSOs for which the data provide significant frequency coverage to analyze the full shapes of the radio spectrum. In section \ref{comparison_sec} we give the details of the criteria used to built the "comparison sample" of normal RL QSOs and we will present the catalog of the 14 used for the radio studies discussed in this section.   

\subsection{Comparison sample}
\label{comparison_sec}

 The "comparison sample" was extracted from the spectroscopically identified non-BAL RL QSOs included in the sample of 36,267  pre-selected sources presented in section \ref{BALselection_sec}. The sample was selected using the same criteria  applied in \cite{DiPompeo_2012}, i.e. searching for each BAL QSOs of our sample, a correspondent non-BAL RL QSOs matched within 20\% SDSS i-band magnitude, 20\% of 1.4 GHz radio flux density, and 10\% of redshift.

\begin{figure} 
\centering 
\includegraphics[width=85mm]{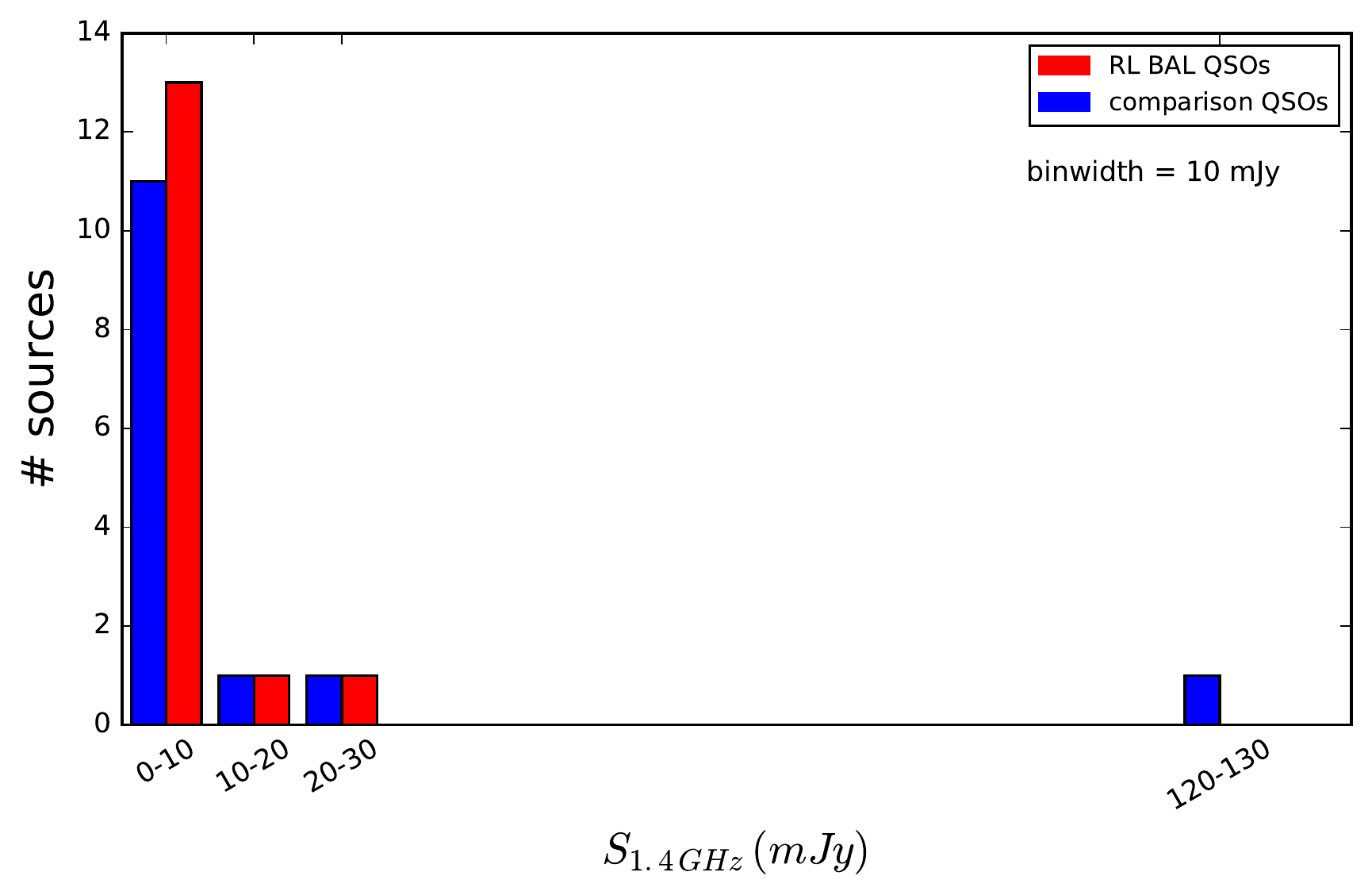}%
\qquad
\qquad 
\includegraphics[width=85mm]{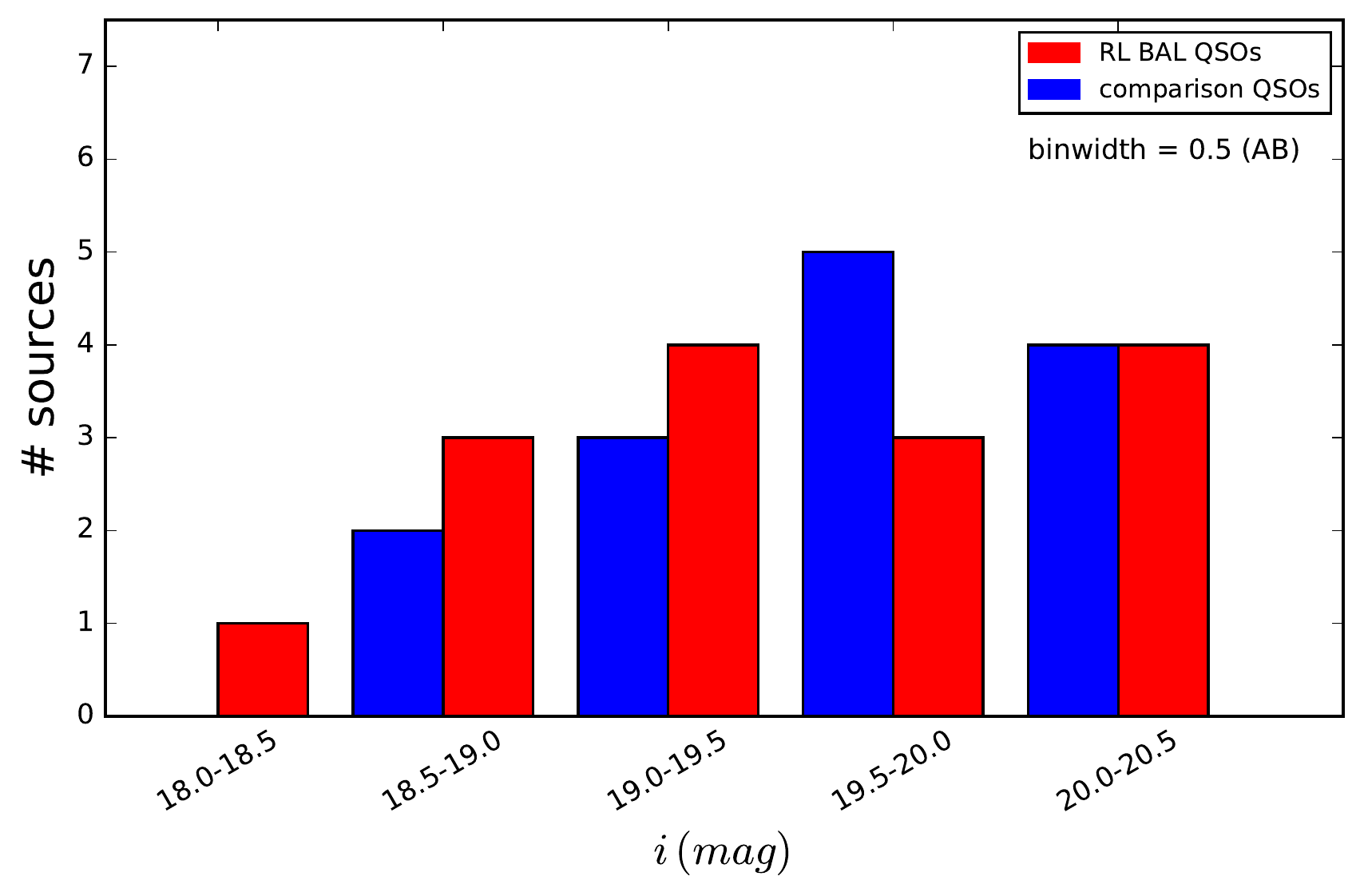} 
\qquad
\qquad 
\includegraphics[width=85mm]{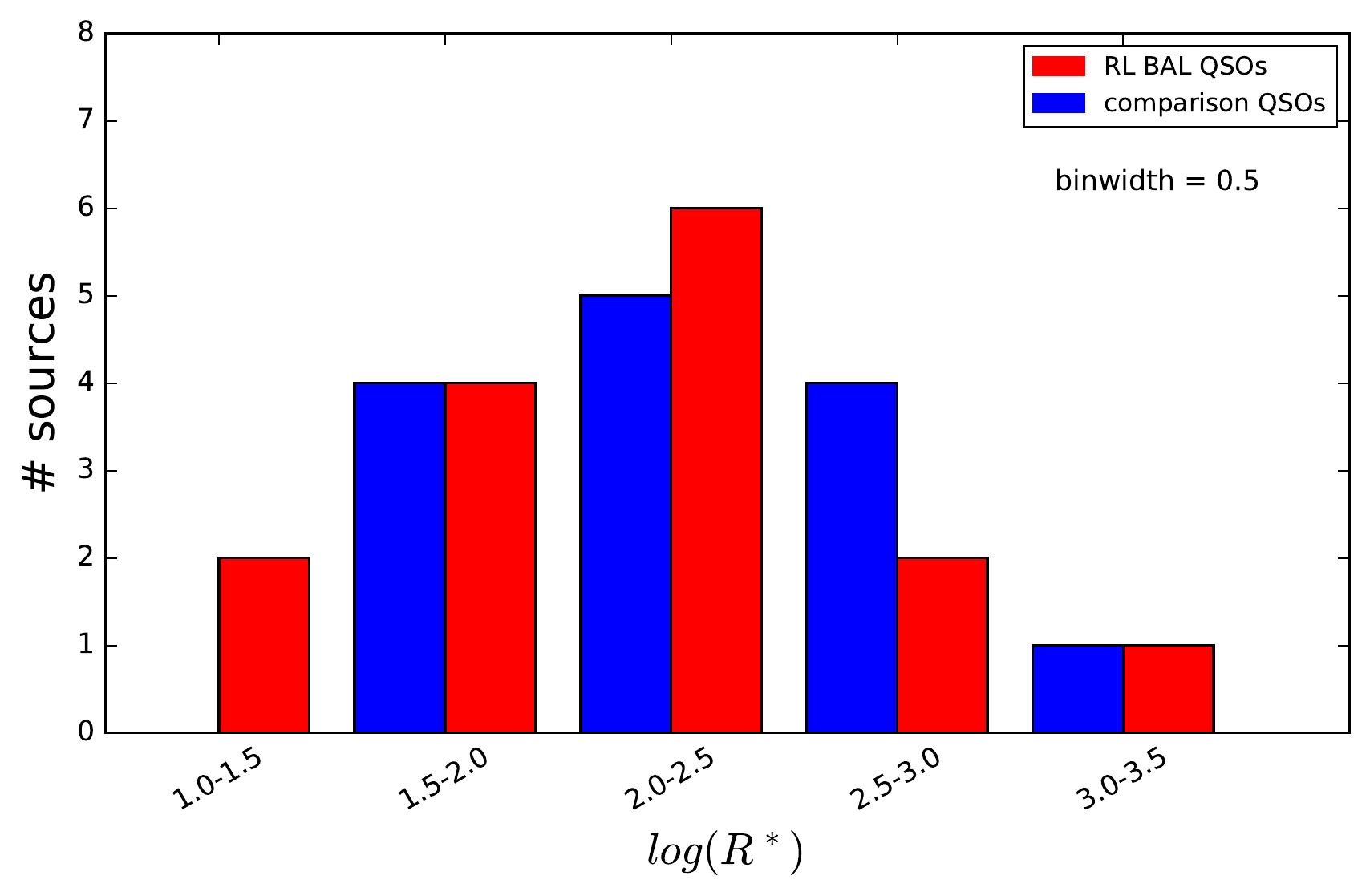} 
\caption{\small  We present the distributions in radio power (FIRST integrated radio flux density), i-magnitude, and radio-loudness $R$ for 15 RL BAL QSOs (red bins) and of 14 normal RL QSOs (blue bins) having $ 3.6 \le z \le 4.8$, as listed in Table 1 and 2. For all these sources we had significant frequency coverage to analyze the full shapes of the radio spectrum. The two samples are well matched in optical magnitude and radio flux density. See section \ref{comparison_sec}}
\label{BALvsComparison}
\end{figure}

Table \ref{Table_comparison} provides a catalog of the 14 non-BAL QSOs of the comparison samples for which the radio-data provide significant frequency coverage to analyze the full shape of the radio spectrum. Fig. \ref{BALvsComparison} show that these 14 non-BAL QSOs and the 15 observed BAL QSOs  are still well matched in redshift, optical magnitude and radio flux density.

 As a whole, the sample of 29 RL quasars (15 BAL and 14 non-BAL) that we observed in radio, extend the samples used in \cite{DiPompeo_2011} and \cite{Bruni_2012}, for the redshift range covered (out to $z=4.8$) and extend to lower optical and radio fluxes. In particular, for their list of 25 radio bright ($S_{1.4\,\rm{GHz}} > 30 \, \rm{mJy}$) RL QSOs with $1.73 \le z \le 3.38$, \cite{Bruni_2012} collected radio flux densities from 1.4 to 43 GHz, including observations with the Effelsberg telescope at the same frequencies (2.6, 4.8, and 8.3 GHz) that we use, and observations at (1.4, 4.86  and 8.46 GHz) with the Very Large Array (VLA) in C configuration.  While for their sample of 74 ($S_{1.4\,\rm{GHz}} > 10 \, mJy$) RL BAL QSOs with $1.5 \le z \le 3.42$,  \cite{DiPompeo_2011} collected flux densities at 4.9 and 8.4 GHz with VLA (in D- and DnC-array configuration) telescope, and completed their observations with the FIRST integrated 1.4 GHz flux densities. Our radio observations including the same, or close, frequencies used in the two cited works,  significantly increase their sample size and therefore determine whether the spectral index distribution differences identified remain robust. This allows us to test not only the orientation properties for a higher redshift population, but also a larger range of radio brightness.

\subsection{Observational campaign}
The observational campaign carried out for this work included both interferometric (JVLA) and single-dish (Effelsberg-100m) observations. In the following, details about observing setup and strategy, as well as data reduction process, are given. See Table \ref{observations} for a summary of observations setups, including observing frequencies and beam sizes for both telescopes.

\begin{table}
 \centering
 \caption{Observing frequencies and beam sizes (half-power beam-width).}
 \label{observations}
 \begin{tabular}{cccc}
 \hline
 Telescope & \multicolumn{1}{c}{Frequency} & \multicolumn{1}{c}{Bandwidth} & \multicolumn{1}{c}{$\theta_{\rm{HPBW}}$}  \\ 
           &  \multicolumn{1}{c}{(GHz)}    &  \multicolumn{1}{c}{(MHz)}    & \multicolumn{1}{c}{(arcsec)}     \\ 
\hline  
Effelsberg-100m & 2.64 &     80  & 265    \\
Effelsberg-100m & 4.85 &    500  & 145    \\
Effelsberg-100m & 8.35 &   1100  &  80    \\
\hline
JVLA(A)     & 1.5  &  1024  &  1.3  \\
JVLA(A)     & 5.5 &  2048  &  0.33  \\
JVLA(A)     & 9.0 &  2048 &   0.20 \\
\hline
\end{tabular}
\end{table}


\subsubsection*{JVLA}

During November 2012, we performed observations at the Jansky Very Large Array observations at 1.5, 5.5, and 9 GHz, with the aim of adding high-sensitivity flux density measurements, and obtain morphological information. All observations were performed in A-configuration, and in dynamic mode, with 1 or 1.5 hours slots. Phase referencing was applied to all sources, and standard flux density calibrators were observed at least once for every slot. A minimum RMS of  $\sim$0.1 mJy was obtained. 
Data were reduced with the latest stable version of the CASA\footnote{http://casa.nrao.edu/index.shtml} software (4.1.0),
making use of customized reduction scripts run on the MPIfR High-Performance Computer cluster (HPC). Flux densities were extracted via bi-dimensional Gaussian fit on the produced maps. Errors were calculated assuming a 5\% uncertainty for the absolute flux density calibration, and quadratically adding it to the map RMS. Given the wide bandwidth available, we split it in two equal intervals, taking as a reference the central frequency, to improve the frequency coverage.

\subsubsection*{Effelsberg-100m telescope}

Observation with the Effelsberg-100m single dish were performed in different runs, initially as a granted-time project in September 2012, and later continued as a filler project, to improve frequency coverage, until May 2015. The cross-scan observing mode was used, at three different frequencies (2.6, 4.8, and 8.3 GHz), increasing the number of sub-scan repetitions depending on the source faintness, and reaching a maximum of 20 minutes observing time per frequency, at which a minimum RMS of $\sim$1 mJy can be reached. Pointing on sources near to target was performed every time the telescope significantly changed the elevation, to compensate for gravitational deformations of the antenna. Data were reduced with the latest version of the TOOLBOX\footnote{https://eff100mwiki.mpifr-bonn.mpg.de/doku.php} software reduction package. Flux density scale was calibrated on well-known sources observed every $\sim$4 hours (3C286, 3C48, 3C295) using the flux densities from \citet{Baars}. Flux densities were extracted via Gaussian fit of the cross-scans. Errors- were calculated assuming a 10\% of uncertainty for the absolute flux density calibration (given the variable weather conditions during the campaign), and adding it via quadratic sum to the cross-scan RMS. Three sigma upper limits are given for non-detections.

\begin{table*}
\caption{Flux densities (mJy) collected in the radio band for the 15 high-redshift BAL QSOs. Measurements at 1.25, 1.75, 5, 6, 8.5, and  9.5 GHz are from the JVLA, while at 2.6, 4.8, and 8.3 GHz are from the Effelsberg-100m single dish. Asterisked values at 1.4 GHz comes from the FIRST survey. In the last three columns, the spectral index, fit type, and peak frequency in the observer's frame (in GHz) are given.}
\begin{center}
\scalebox{0.88}{
\begin{tabular}{cccccccccccccccc}
\hline
Source 		&  $S_{1.25}$ 		& $S_{1.4}$ 		& $S_{1.75}$ 		& $S_{2.6}$		& $S_{4.8}$ 		&	$S_{5}$ 		& $S_{6}$  		&	$S_{8.3}$ &  $S_{8.5}$ 		& $S_{9.5}$ 		& $\alpha$ 		& Fit & Peak 	\\ 
\hline
0000+00	    	&   2.0$\pm$0.1	&  -	  	&	1.4$\pm$0.2  	& 	<11	&	 -		&	0.8$\pm$0.1  	&	0.7$\pm$0.1  	&	-	&    0.8$\pm$0.1  	&	0.7$\pm$0.1  	&  -0.00$\pm$0.29	&  L	&    - 	\\		
0747+13	    	&	-   	&  8.0$\pm$0.8*   	&	-    	& 	 12$\pm$4 	&	13$\pm$2		&	-    	&   -   		&  6$\pm$1    	&	-    	&   -   		&  -1.41$\pm$0.37	&  P	&    3.14 	\\
1006+46     	&	-   	&  6.3$\pm$0.6*   	&	-    	& 	 12$\pm$1 	&	 8$\pm$1		&	-    	&   -   		&  5$\pm$1    	&	-    	&   -   		&  -0.86$\pm$0.38	&  P	&    3.02 	\\
1023+55     	&	4.1$\pm$0.3	&  - 	  	&	4.8$\pm$0.4  	& 	<23  	&	 4$\pm$1		&	2.1$\pm$0.1  	&	1.4$\pm$0.1  	&	-    &	1.6$\pm$0.1  	&	1.5$\pm$0.1  	&  -0.51$\pm$0.13	&  P	&    1.99 	\\
1036+50	    	&	-    	&  9.2$\pm$1.0*   	&	-    	& 	  7$\pm$1 	&	 5$\pm$1		&	-    	&   -   		&  3$\pm$1    	&	-    	&   -   		&  -0.93$\pm$0.63	&  P	&   0.56 	\\
1109+19	    	&	7.2$\pm$0.5	&  -      		&	5.7$\pm$0.3  	& 	  6$\pm$1 	&     	 -		&	2.8$\pm$0.1  	&	2.4$\pm$0.1  	&	-    &	1.7$\pm$0.1  	&	1.5$\pm$0.1  	&  -0.94$\pm$0.11	&  P	&   0.29 	\\
1110+43     	&	-    	&  1.2$\pm$0.2*   	&	-    	& 	  -  		&	 3$\pm$1		&	-    	&   -   		&  3$\pm$1    	&	-    	&   -   		&  0.00$\pm$0.77	&  P	&    6.31 	\\
1129+13	    	&	7.0$\pm$1.5	&  -	  	&	2.1$\pm$0.2  	& 	  -  		&	 -  		&	1.0$\pm$0.1  	&	0.6$\pm$0.1  	&	-    &	0.5$\pm$0.1  	&	0.4$\pm$0.1  	&  -1.31$\pm$0.36	&  P	&   0.97 	\\	     
1157+22	    	&	2.4$\pm$0.1	&  -	  	&	2.1$\pm$0.1  	& 	  -		&	 -  		&	2.2$\pm$0.1  	&	1.9$\pm$0.1  	&	-    &	1.3$\pm$0.1  	&	1.1$\pm$0.1  	&  -0.99$\pm$0.15	&  P	&    3.05 	\\	     
1332+00     	&	2.7$\pm$0.2	&  -	  	&	1.8$\pm$0.1  	& 	  -  		&	 -  		&	0.7$\pm$0.1  	&	0.7$\pm$0.1  	&	-    &	0.5$\pm$0.1  	&	0.5$\pm$0.1  	&  -0.63$\pm$0.40	&  L	&    - 	\\	     
1344+62	    	&	2.0$\pm$0.3	&  -	  	&	2.3$\pm$0.2  	& 	 <6	&	 -  		&	0.9$\pm$0.1  	&	0.7$\pm$0.1  	&	-    &	-    	&	-    	&  -1.38$\pm$0.30	&  P	&    1.77 	\\	     
1348+17     	&	2.0$\pm$0.3	&  -	  	&	1.7$\pm$0.1  	& 	  4$\pm$1 	&	 -  		&	3.1$\pm$0.2  	&	3.0$\pm$0.2  	&	-    &	2.6$\pm$0.1  	&	2.5$\pm$0.2  	&  -0.33$\pm$0.12	&  P	&    4.85 	\\
1506+53     	&	-       	& 14.6$\pm$1.5*   	&	-    	& 	 14$\pm$3 	&	 -  		&	-    	&   -   		&	  7$\pm$2  &	-    	&   -   		&  -0.60$\pm$0.58	&  P	&    1.70 	\\	     
1511+25     	&	-    	&  1.2$\pm$0.2*   	&	-    	& 	  -  		&       <12 	&	-    	&   -   		&  8$\pm$2    	&	-    	&   -   		&   1.07$\pm$0.49	&  L	&    - 	\\
1659+21	    	&  29.1$\pm$1.5	&  -	  	&    24.6$\pm$1.3  	& 	 20$\pm$2 	&	16$\pm$4		&      12.1$\pm$0.6  	&	10.2$\pm$0.5 	&	-    	&    7.9$\pm$0.5  	&	6.9$\pm$0.4  	&  -0.80$\pm$0.13	&  P	&   0.16 	\\
\hline
\end{tabular}
}
\end{center}
\label{BAL-flux}
\end{table*}%

\begin{table*}
\caption{Flux densities (mJy) in the radio band collected for the 14 high-redshift non-BAL QSOs. Measurements at 1.25, 1.75, 5, 6, 8.5, and 9.5 GHz are from the JVLA, while at 2.6, 4.8, and 8.3 GHz are from the Effelsberg-100m single dish. Asterisked values at 1.4 GHz comes from the FIRST survey. In the last three columns, the spectral index, fit type, and peak frequency in the observer's frame (in GHz) are given.}
\begin{center}
\scalebox{0.88}{
\begin{tabular}{cccccccccccccccc}
\hline
Source 	&  $S_{1.25}$ 		& $S_{1.4}$ 		& $S_{1.75}$ 		& $S_{2.6}$		& $S_{4.8}$ 		&	$S_{5}$ 		& $S_{6}$  	&	$S_{8.3}$ &  $S_{8.5}$ 		& $S_{9.5}$ 		& $\alpha$ 		& Fit 			& Peak 	\\ 
\hline
0300+00	   &	6.4$\pm$0.4  	&	   -   	&     7.5$\pm$0.4  	&	  6$\pm$1		&	-   		&    4.3$\pm$0.2  	&	3.5$\pm$0.2 	& 	-     	&  1.4$\pm$0.1  	& 1.3$\pm$0.1  	&  -2.11$\pm$0.14  	&  P  &        	2.13  \\
0833+09	   &	-    	&	 126$\pm$12*   	& 	   -  	&	106$\pm$11	&	88$\pm$9   	&	-    	&   -  		&   77$\pm$14    	&  -    		&   -   		&  -0.24$\pm$0.34	&  L  &	       -     \\
0840+34	   &	-    	&	13.6$\pm$0.1*  	& 	   -  	&	 15$\pm$2	&	31$\pm$6   	&	-    	&   -  		&    8$\pm$1     	&  -    		&   -   		&  -2.47$\pm$0.38	&  P  &	       4.26  \\
0901+10	   &	3.4$\pm$0.5  	&	  -    	&     2.0$\pm$0.3  	&	  -		&	-   		&	0.8$\pm$0.1  	&	0.6$\pm$0.1 	& 	-     	&  0.4$\pm$0.1  	&	0.3$\pm$0.1  	&  -1.31$\pm$0.46	&  L  &	       -     \\
0918+06	   &	-    	&	26.5$\pm$2.6*  	& 	   -  	&	 43$\pm$4	&	36$\pm$4   	&	-    	&   -  		&   23$\pm$4     	&  -    		&   -   		&  -0.82$\pm$0.34	&  P  &	       3.15  \\
1017+34	   &	-    	&	 2.6$\pm$0.3*  	& 	   -  	&	  7$\pm$1		&	11$\pm$1   	&	-    	&   -  		&   -     		&  -    		&   -   		&   0.74$\pm$0.28	&  P  &	       5.98  \\
1102+53	   &	-    	&	 5.6$\pm$0.6*  	& 	   -  	&	  6$\pm$1		&	 6$\pm$1   	&	-    	&   -  		&    6$\pm$1     	&  -    		&   -   		&  -0.00$\pm$0.38	&  L  &	       -  \\
1125+57	   &	2.3$\pm$0.2  	&	 -    	& 	2.0$\pm$0.1  	&	  -		&	-   		&	0.8$\pm$0.1  	&	0.6$\pm$0.1 	& 	-     	&  0.5$\pm$0.1  	&	0.5$\pm$0.1  	&  -0.89$\pm$0.38	&  P  &	      0.84  \\
1150+42	   &	3.0$\pm$0.2  	&	 -    	&	2.0$\pm$0.2  	&	  -		&	-   		&	0.5$\pm$0.1  	&	0.4$\pm$0.1 	& 	-     	& <0.3    	& <0.3    	&  -1.73$\pm$0.88	&  L  &	       -     \\
1249+15	   &	2.4$\pm$0.3  	&	 -    	&	1.2$\pm$0.1  	&	  -		&	-   		&	0.6$\pm$0.1  	&	0.5$\pm$0.1 	& 	-     	&  0.3$\pm$0.1  	& <0.3    	&  -1.31$\pm$0.61	&  P  &	       3.10  \\
1311+22	   &	6.3$\pm$0.4  	&	 -    	&	5.4$\pm$0.3  	&	  7$\pm$2 	&	-   		&	-    	&	-   	& 	-     	&  1.4$\pm$0.1  	&	1.3$\pm$0.1  	&  -1.36$\pm$0.48	&  P  &	      0.71  \\
1423+39	   &	6.6$\pm$0.4  	&	 -    	&    10.4$\pm$0.6  	&	<21	&	-   		&   11.4$\pm$0.8  	&	9.5$\pm$0.5 	& 	-     	&  7.4$\pm$0.4  	&	6.8$\pm$0.3  	&  -0.81$\pm$0.14	&  P  &	       3.36  \\
1446+60	   &	3.0$\pm$0.3  	&	 -    	&	2.4$\pm$0.2  	&	 -	&	-   		&	1.4$\pm$0.1  	&	1.3$\pm$0.1 	& 	-     	&  0.7$\pm$0.1  	&	0.6$\pm$0.1  	&  -1.31$\pm$0.26	&  P  &	       1.28  \\
1611+08	   &   20.1$\pm$1.2  	&	 -    	&    16.6$\pm$0.9  	&	  -		&	-   		&   14.2$\pm$0.7  	& 14.2$\pm$0.7 	& 	-     	& 11.5$\pm$0.6  	& 10.8$\pm$0.5  	&  -0.40$\pm$0.12  	&  P  &	      0.88  \\
\hline
\end{tabular}
}
\end{center}
\label{comparison-flux}
\end{table*}%

\section{Observational results}

The collected flux densities are reported in Tab. \ref{BAL-flux} and \ref{comparison-flux}. In total, 20 sources were observed with the JVLA, and 26 with the Effelsberg-100m. When no detection at 1.5 GHz were available from our campaign, we used the measurement at 1.4 GHz from the FIRST survey. 

Despite observations at the Effelsberg-100m telescope have been carried out on a 3-years time window, the obtained SED do not show significant changes along frequencies, that might suggest a strong variability. Thus, we can safely combine data from this campaign to obtain the desired frequency coverage.

\subsection{Morphology}

All the observed sources are unresolved both with the Effelsberg-100m and the JVLA in A configuration. Considering the angular resolution of the latter for the 8 BAL and 8 non-BAL QSOs detected at 9 GHz (0.2 arcsec), we obtain an upper limit for the projected linear size of 1.4 kpc at the mean redshift of 4.02 and 3.91 for our BAL and non-BAL samples. This is compatible with the typical linear size of High Frequency Peakers (HFP, 0.01-0.5 kpc, \citealt{Dallacasa} and also GPS sources (0.5-5 kpc, \citealt{ODea}). Further Very Long Baselines Interferometry (VLBI) observations would be required to resolve sources at these redshifts.

\begin{figure*}
\begin{center}
\includegraphics[width=17cm]{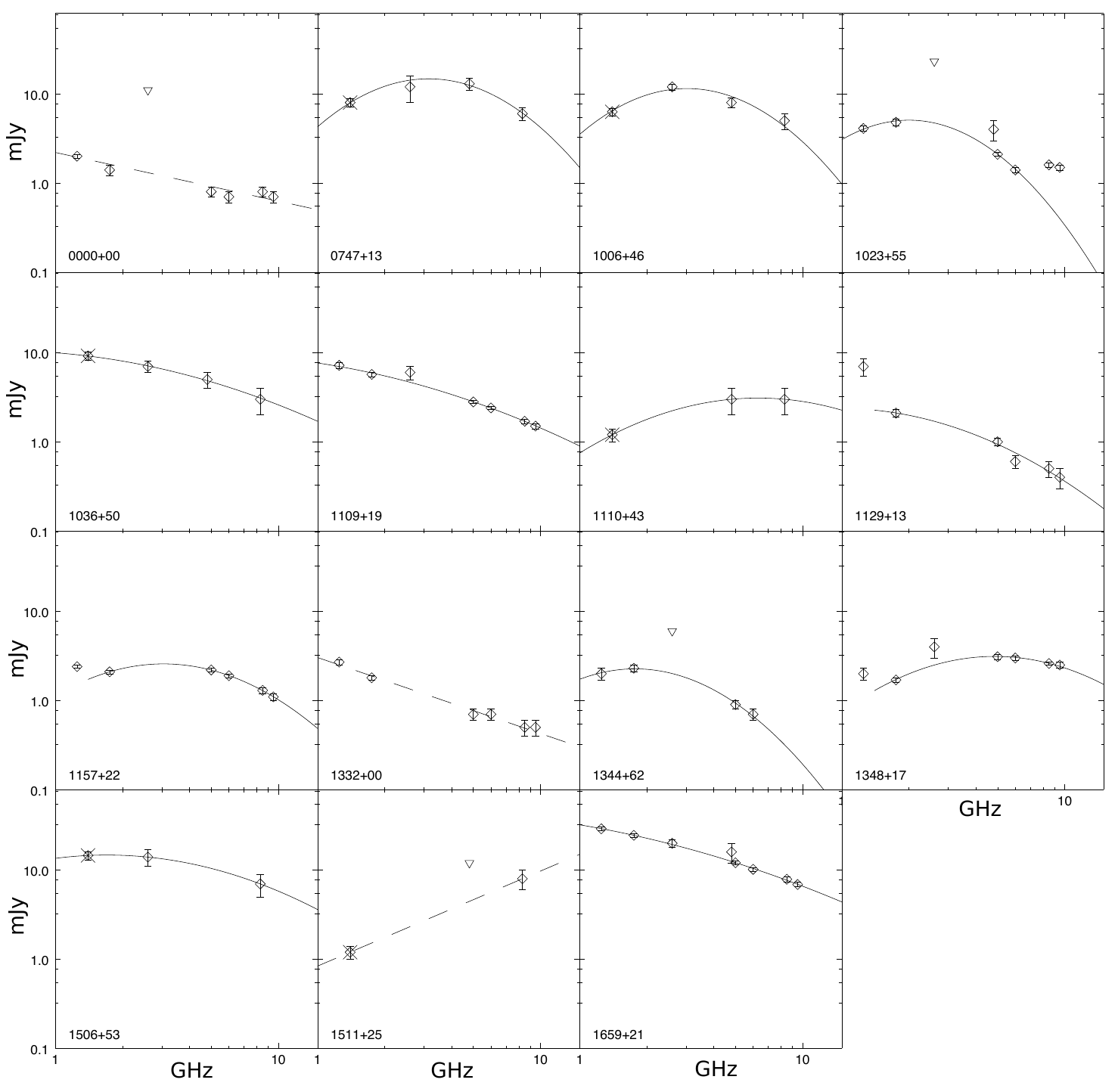}
\caption{SEDs of the 15 high-redshift BAL QSOs studied here (x-axis: GHz, y-axis: mJy). Flux densities from the FIRST catalogue, at 1.4 GHz, are plotted as crosses. Triangles are 3$\sigma$ upper limits.}
\label{BALs}
\end{center}
\end{figure*}

\begin{figure*}
\begin{center}
\includegraphics[width=17cm]{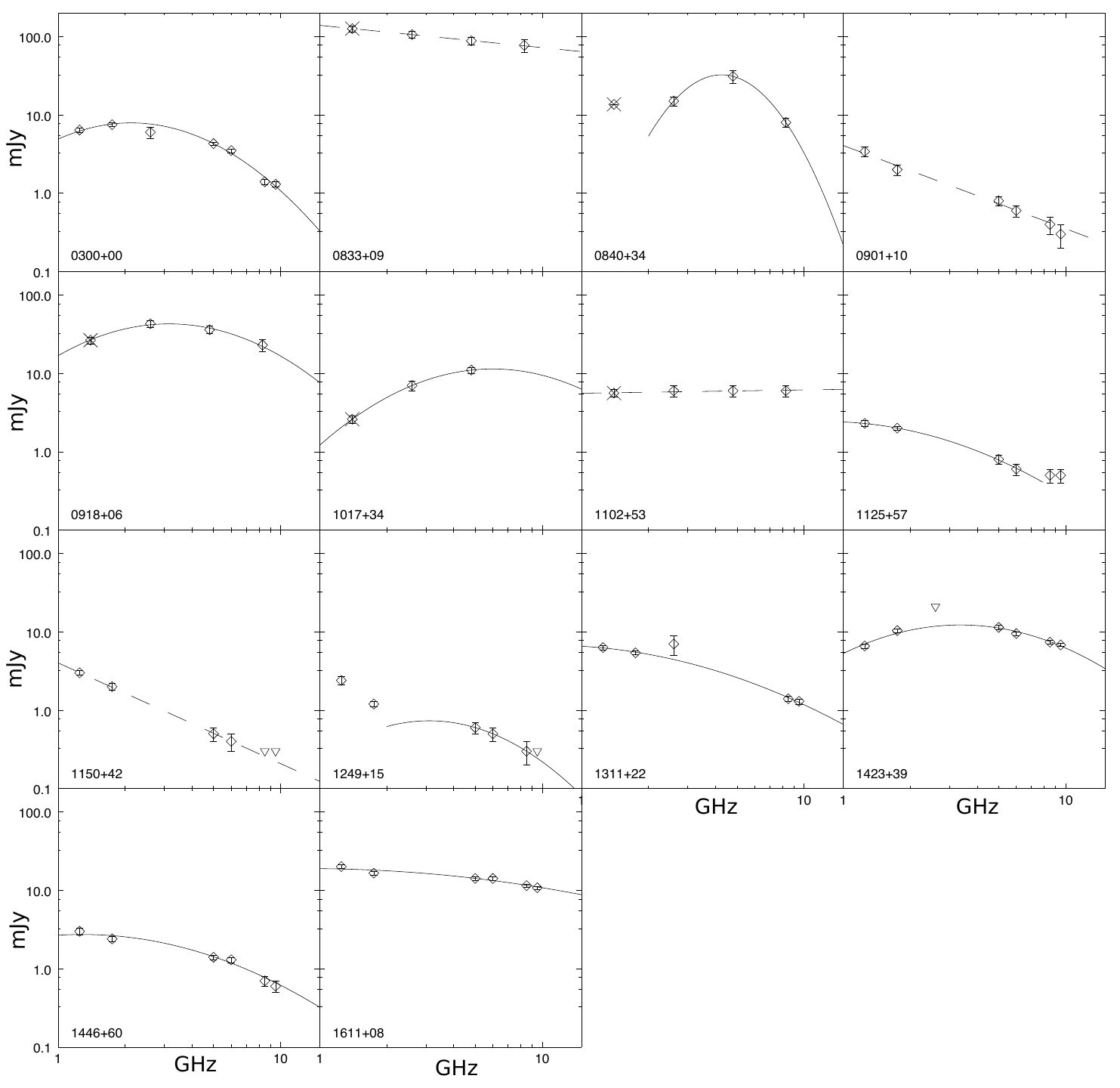}
\caption{SEDs of the 14 comparison non-BAL QSOs (x-axis: GHz, y-axis: mJy). Flux densities from the FIRST catalogue, at 1.4 GHz, are plotted as crosses. Triangles are 3$\sigma$ upper limits.}
\label{comparison}
\end{center}
\end{figure*}

\subsection{SED shape and spectral index}

In the following we analyze the SEDs in order to get information about the orientation and age of our high-redshift sample.

\subsubsection*{Synchrotron peak frequencies} 

The collected flux densities between 1.25 and 9.5 GHz allow us to reconstruct the SED of our objects in a reasonable way. When no measurement from our campaign were available at 1.5 GHz, we used data from the FIRST survey. In Fig. \ref{BALs} and \ref{comparison} the SEDs for each source are presented. We can get an indication of the presence of a peak in order to obtain the fraction of young radio sources in our samples. Indeed, GPS sources at lower redshifts are usually identified as objects peaking in the range 1 GHz$\le\nu_{peak}\le$8 GHz (\citealt{ODea}) while HFP can peak at frequencies $>$8 GHz \citep{Dallacasa}. Both are generally interpreted as young radio sources, in a evolutionary track towards CSS sources, peaking at hundred of MHz (\citealt{Saikia_1988}; \citealt{Fanti_1990}), as the emitting plasma adiabatically expands. Considering the redshift range of our sample (mean redshift $\sim$4), the peaking interval of GPS and HFP translates - in the observer's frame - into an interval between 0.2 GHz and 1.6 GHz for the first (outside our observing window) and >1.6 GHz for the latter. 

We performed SED fitting using two simple functions: a power-law (`L' hereafter) describing the optically thin part of the synchrotron emission, and a parabola in logarithmic scale (`P' hereafter) as a simplified version of the peaked synchrotron emission, including the optically thin/thick parts. The aim was to determine, via chi-squared minimization, which among our sources present a peak in the explored frequency range. Only for one source, 1511+25, we have two points for the SED, not allowing a parabolic fit. Thus, for this source we can not discard the presence of a peak, however given the  clearly inverted spectrum between 1.4 and 8.3 GHz, and the upper limit at 4.8 GHz, a peak in this range is improbable. For some sources we excluded from the fit the measurements that clearly belong to a second synchrotron component, at at higher or lower frequency range with respect to the fitted one. These were: 1.4 GHz for 0840+34, 1129+13, 1157+22, 1348+17; 1.25 GHz and 1.75 GHz for 1249+15; 8.5 GHz and 9.5 GHz for 1023+55, 1125+57. The results are reported in Tables \ref{BAL-flux} and \ref{comparison-flux}, including the observer's frame peak frequency, when present.

In total, 12 out of 15 BAL QSOs (80\%) and 10 out of 14 non-BAL QSOs (71\%) could be fitted with a P, and thus show hints of a peak, not suggesting a clear difference between the two samples. 
A selection effect, due to the requirement that sources are detected in FIRST despite the large redshift, could also have a role in selecting sources peaking in the 1-50 GHz (rest-frame) range. Nevertheless, the use of a comparison sample - selected in the same way - allows us to draw statistically significant conclusions about the differences between BAL and non-BAL QSOs.


\subsubsection*{Spectral indices}
The spectral index of the synchrotron spectrum (defined here as $\alpha$ in the expression $S=\nu^\alpha$) can be a useful orientation indicator \citep{Orr_1982} . This has been used in previous works to characterize the orientation angle distribution of RL BAL QSOs (\citealt{Montenegro}, \citealt{DiPompeo_2011}, \citealt{Bruni_2012}, \citealt{DiPompeo_2012}) finding only a mildly preferred orientation with respect to non-BAL QSOs. We repeat here the same analysis, to test the orientation scenario at high redshifts. 
For the calculation, we used the flux density measurements at 5 and 8.5 GHz from the JVLA, or the ones 4.8 and 8.3 GHz from the Effelsberg-100m dish when the previous were not available. In this way, we have an estimate as homogeneous as possible for the different sources, and consider frequencies at the right side of the peak, when present. The only exceptions are the two sources presenting an inverted spectrum in the observed frequency range: BAL QSO 1511+25, and non-BAL QSO 1017+34. In this case we are looking at the self-absorbed part of the synchrotron spectrum. The obtained values for the two samples are given in Table \ref{BAL-flux} and \ref{comparison-flux}.

Considering only the objects that do not present an inverted spectrum (i.e. $\alpha<0$, 14 BAL and 13 non-BAL QSOs) 12/14 BAL QSOs (86\%) are steep ($\alpha<-0.5$), while 10/13 non-BAL QSOs (77\%) are steep. A Kolmogorov-Smirnov (K-S) test on the two spectral index distributions results in a \emph{p}=0.40, that does not allow us to exclude the null-hypothesis that the two samples of values are drawn from the same parent distribution. Also considering the whole spectral index distributions from Tables \ref{BAL-flux} and \ref{comparison-flux} for the K-S test, we obtain a \emph{p}=0.45, confirming the previous result.

\section{Broadband colors of high-z RL BAL QSOs}
\label{secBrooadCol}
The continuum emission differences between BALQSOs and non-BALQSOs can be investigated by studying the differences between their broadband colors, since they can give an indication of the photometric spectral index. In this section we investigate the radio loudness, the broadband optical (subsection \ref{sec_optical}) and infrared (subsection \ref{sec_infrared}) colors of RL BAL QSOs in the highest redshift range for which the \ion{C}{iv} $\,$  absorption troughs is still visible in the optical spectra. For this purpose we use all the 22 RL BAL QSOs selected in section 2 and listed in Table \ref{BALTable}. We compare them with normal RL QSOs in the same range of redshift  ($ 3.6 \le  z \le 4.8$) and detected both in SDSS DR7 and FIRST. This comparison sample consists of 99 QSOs from the SDSS DR7 QSO catalogue and it is complemented by another 14 spectroscopical confirmed QSOs selected using a 97\% complete selection strategy described in \cite{Tuccillo} and based on the use of an artificial neural network to select quasar candidates within sources without spectra in SDSS. This way resulting a total of 113 RL QSOs. We note that we excluded from the comparison sample 5 QSOs showing broad absorption in the \ion{o}{iv} $\,$ emission-line only (and not in \ion{C}{iv} $\,$) in order to reduce any bias in the comparison between BALs and non-BALs quasars. 
  
\begin{figure} 
\centering 
\includegraphics[width=85mm]{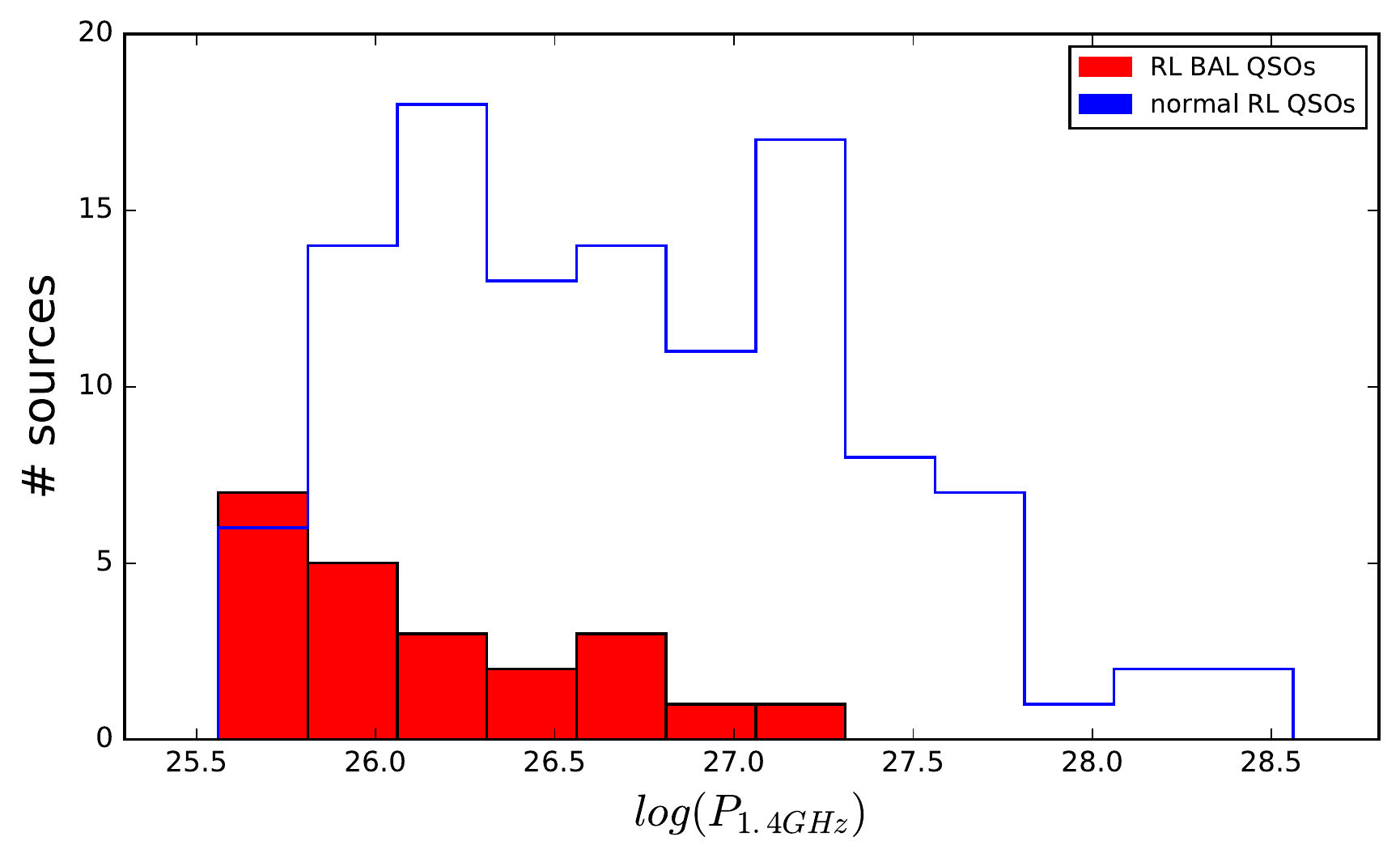}%
\qquad
\qquad 
\includegraphics[width=85mm]{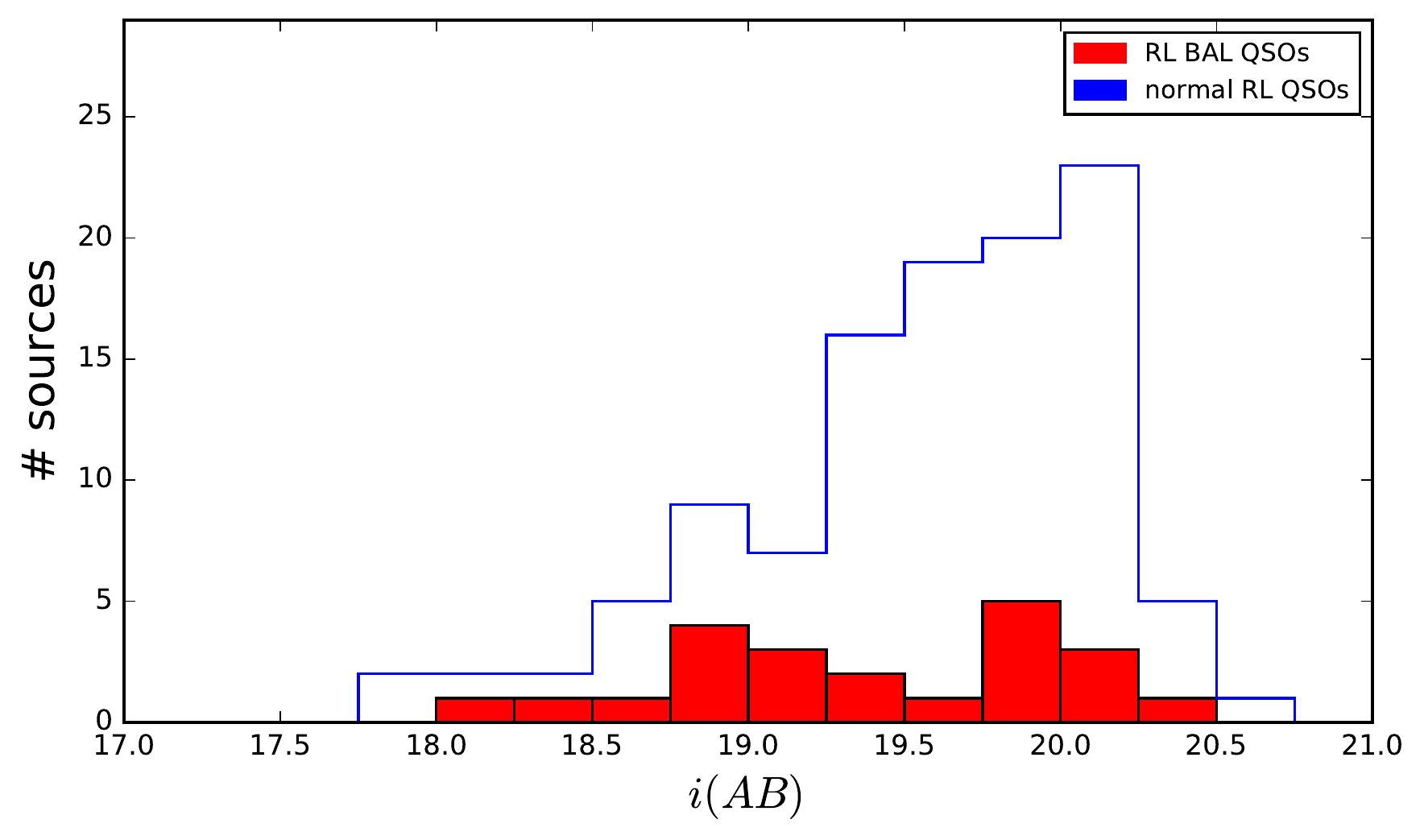} 
\qquad
\qquad 
\includegraphics[width=85mm]{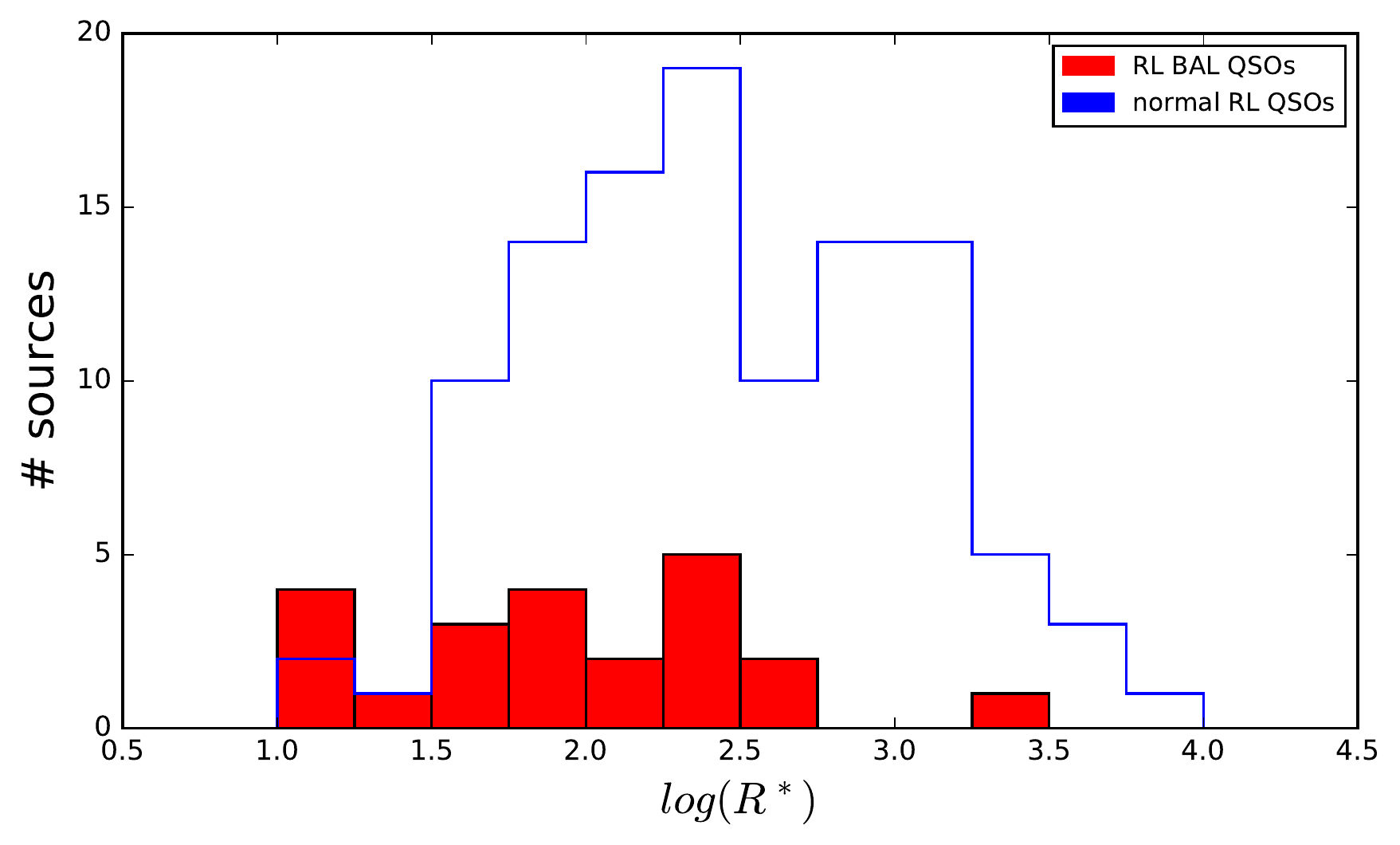} 
\caption{\small Distribution in radio power, i - magnitude, and radio-loudness $R$, for our sample of 22 $ 3.6 \le z \le 4.8$ RL BAL QSOs (red bins) and a sample of 113 normal RL QSOs (blue bins) matched in redshift. We used these samples to compare the differences between broadband optical and NIR colors of RL BAL and non-BAL RL QSOs at high-z. See section \ref{secBrooadCol}.}
\label{distributionsBALsandNon}
\end{figure}

The distribution of the radio power, the i-magnitude and the radio-loudness $R$, for the two samples of 22 BAL and the 113 non-BAL samples is showed in Fig. \ref{distributionsBALsandNon}. All the QSOs of our sample satisfy our criteria to be detected in FIRST ($S_{1.4\,\rm{GHz}} > 1\, \rm{mJy}$) and in SDSS DR7, therefore there is no a-priori reason for which BAL QSOs should be radio- and optical- distributed differently than normal QSOs. In this sense the fact that our BAL sample tends to have lower radio-power, is consistent with the tendency for strongly radio-loud quasars to lack of BAL QSOs (\citealt{Becker_2001}; \citealt{Richards_2011}).


\subsection{Broadband optical colors}
\label{sec_optical} 

\begin{figure} 
\centering 
\includegraphics[width=80mm]{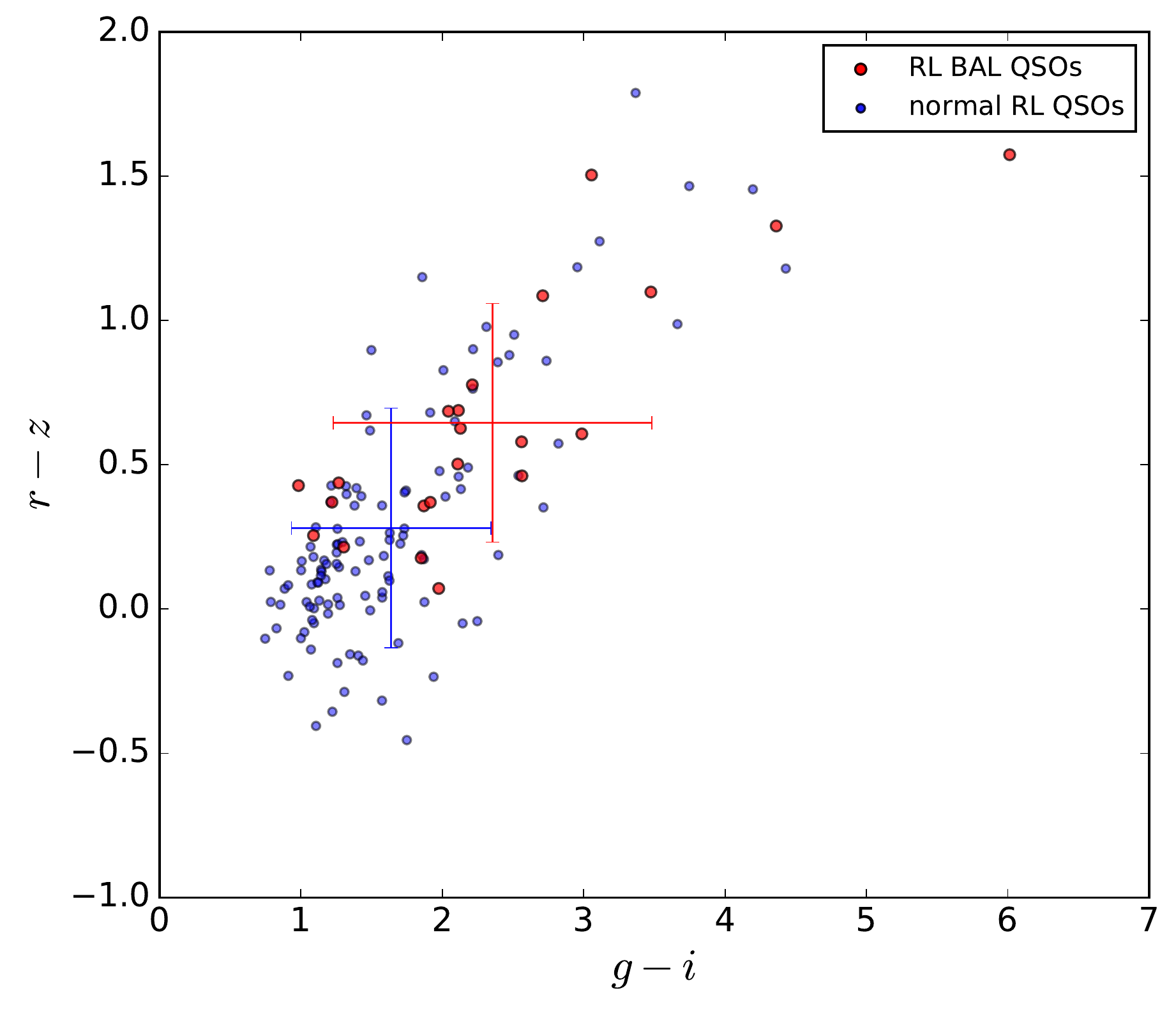} 
\qquad
\qquad 
\includegraphics[width=80mm]{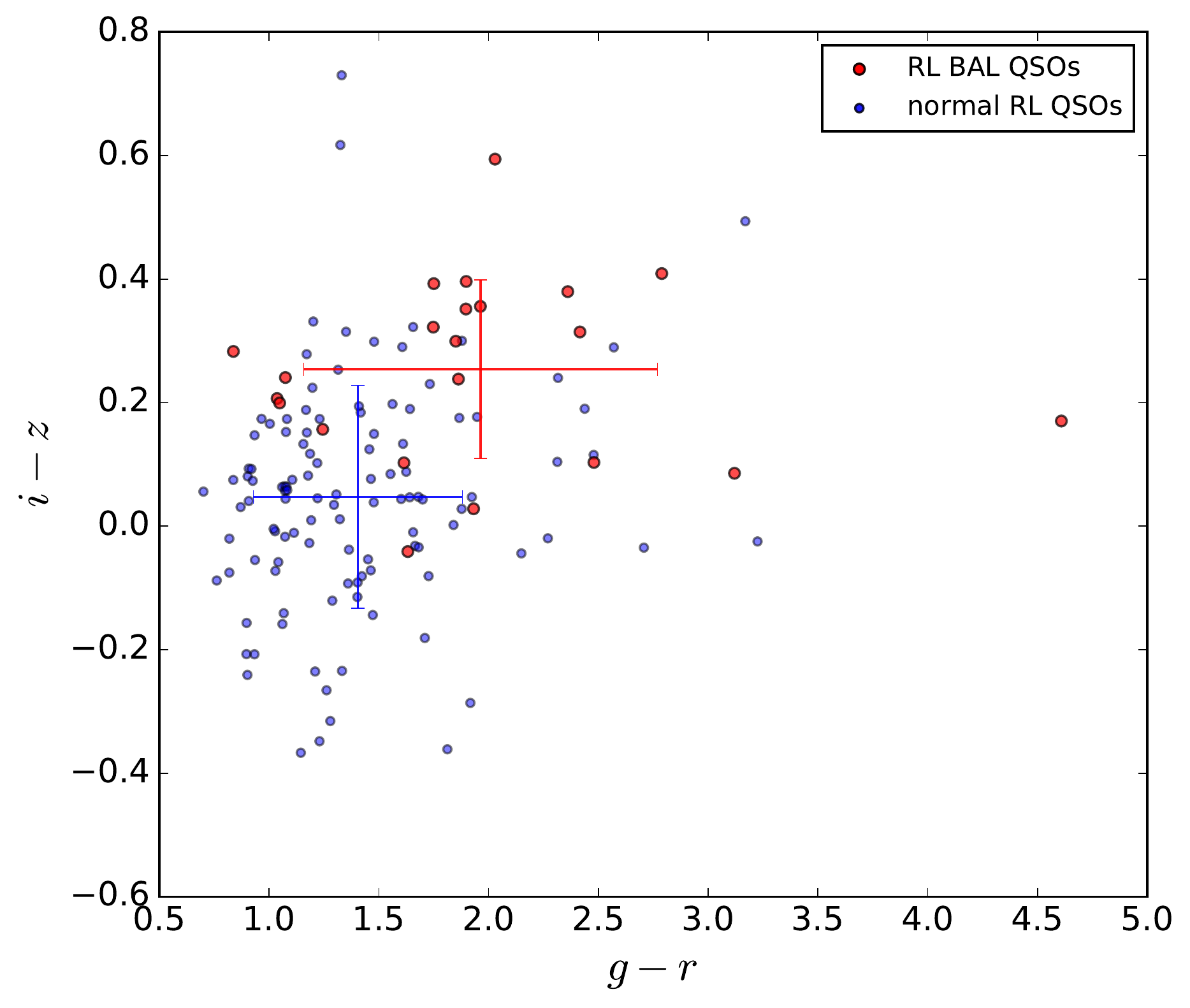}
\caption{\small  Two-color diagrams presenting the SDSS colors of 22 high-z RL BAL (red dots) and 113 normal RL QSOs (blue dots) in the same range of redshift. We show the mean colors of the BAL sample as a red cross and the mean colors of the normal QSOs as a blue cross, both with 1-$\sigma$ error bar. See section \ref{sec_optical}} 
\label{colorColors}
\end{figure}

In Fig. \ref{colorColors} we show SDSS color-color diagrams of our samples of 22 BAL and 113 non-BAL QSO. All the colors have been corrected for galactic extinction using the maps of \cite{Schlegel_1998}. In the plots the large crosses with 1-$\sigma$ error bar, represent the location of the mean for each of the plotted samples,  respectively red for BAL and blue for normal QSOs. BALQSOs tend to occupy redder sub-regions of the color-color space determined by the distribution of the parent population of quasars. This trend is analyzed with a t-test to compare the means of all the SDSS colors for the BALs and non BALs samples. In Table \ref{t_test} we report only the colors for which the t-test rejects the null-hypothesis that the color-vectors come from normal-distributed samples having equal means, and therefore suggesting a significant difference in those values. The t-test is performed in the general conditions of not assuming equal variances (Behrens-Fisher problem) and the number of degrees of freedom $df$ is given by the Satterthwaite approximation.

\begin{table}
\centering
\label{t_test}
\caption{t-test on SDSS colors}
\renewcommand\tabcolsep{4pt}
\scalebox{0.90}{
\begin{tabular}{c c c c c c c c}
\hline \hline
\multicolumn{1}{c}{Color} &
\multicolumn{1}{c}{Population} &
\multicolumn{1}{c}{Mean} &
\multicolumn{1}{c}{$\sigma$} &
\multicolumn{1}{c}{Median} &
\multicolumn{1}{c}{t} &
\multicolumn{1}{c}{df} &
\multicolumn{1}{c}{p ($\times 10^{-2}$)} \\
\multicolumn{1}{c}{(1)} &
\multicolumn{1}{c}{(2)} &
\multicolumn{1}{c}{(3)} &
\multicolumn{1}{c}{(4)} &
\multicolumn{1}{c}{(5)} &
\multicolumn{1}{c}{(6)} &
\multicolumn{1}{c}{(7)} &
\multicolumn{1}{c}{(8)} \\
\hline
\multirow{2}{*}{(u-i)}   & BAL         & 5.13	& 1.02	& 5.21 & \multirow{2}{*}{2.3}  & \multirow{2}{*}{32.2} & \multirow{2}{*}{3.00}  \\
                                   & non BAL  & 4.58	& 1.14	& 4.40 &  \\
\hline
\multirow{2}{*}{(u-z)}   & BAL         & 5.38	& 1.04	& 5.47 & \multirow{2}{*}{3.0}  & \multirow{2}{*}{32.1} & \multirow{2}{*}{0.46}  \\
                                   & non BAL  & 4.62	& 1.17	& 4.48 &  \\
\hline
\multirow{2}{*}{(g-r)}   & BAL         & 1.96	& 0.82	& 1.87 & \multirow{2}{*}{3.0}  & \multirow{2}{*}{23.8} & \multirow{2}{*}{0.55}  \\
                                   & non BAL  & 1.41	& 0.48	& 1.31 &  \\
\hline
\multirow{2}{*}{(g-i)}   & BAL         & 2.36	& 1.15	& 2.11 & \multirow{2}{*}{2.8}  & \multirow{2}{*}{24.1} & \multirow{2}{*}{1.00}  \\
                                   & non BAL  & 1.64	& 0.71	& 1.43 &  \\
\hline
\multirow{2}{*}{(g-z)}   & BAL         & 2.61	& 1.15	& 2.45 & \multirow{2}{*}{3.6}  & \multirow{2}{*}{24.8} & \multirow{2}{*}{0.14}  \\
                                   & non BAL  & 1.69	& 0.77	& 1.46 &  \\
\hline
\multirow{2}{*}{(r-z)}   & BAL         & 0.64	& 0.42	& 0.54 & \multirow{2}{*}{3.7}  & \multirow{2}{*}{29.3} & \multirow{2}{*}{0.08}  \\
                                   & non BAL  & 0.28	& 0.41	& 0.18 &  \\
\hline
\multirow{2}{*}{(i-z)}   & BAL         & 0.25	& 0.15	& 0.26 & \multirow{2}{*}{5.7}  & \multirow{2}{*}{34.3} & \multirow{2}{*}{0.00}  \\
                                   & non BAL  & 0.05	& 0.18	& 0.05 &  \\
\hline \hline                                                                     
\end{tabular}
}

\medskip
\begin{flushleft}
\small  \textit{The columns give the following: (1) SDSS color; (2) subsample considered; (3) mean of the color; (4) standard deviation; (5) median of the color; (6) statistic $t$; (7) associated degrees of freedom to the t-test; (8) the statistic p associated to the t-test, as usually, the null hypothesis of equal mean is rejected for values < 0.05}
\end{flushleft}
\end{table}

\begin{figure} 
\centering 
\includegraphics[width=87mm]{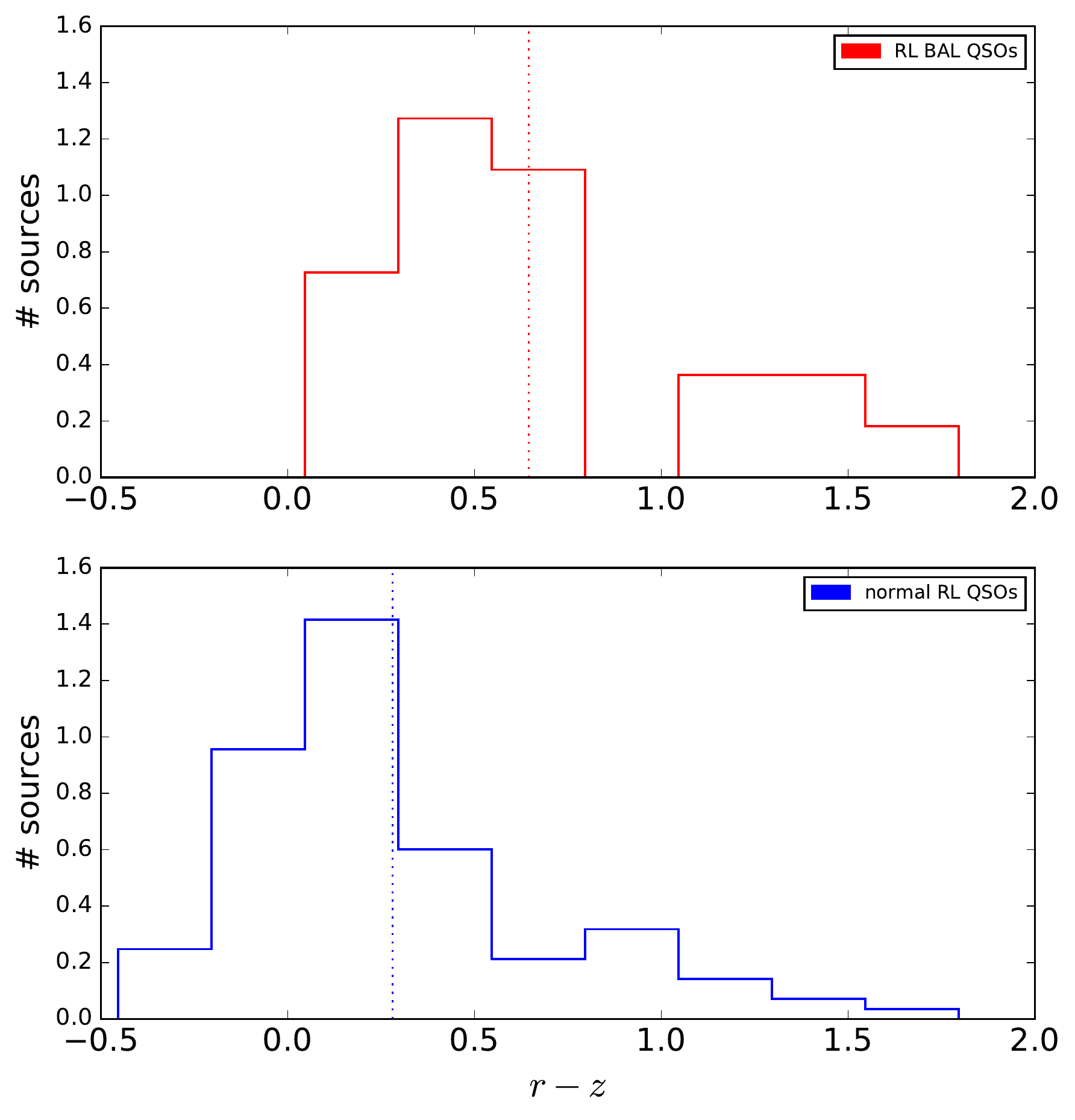} 
\caption{\small  Comparison of the normalized distribution of the (r-z) optical colors of high-z RL BAL and non-BAL QSOs, as discussed in section \ref{sec_optical}. The mean of the BAL QSOs colors is indicated by red a dashed lines, while for the normal RL QSOs is indicated by blue dashed lines.} 
\label{distributionColors2}
\end{figure}

\begin{figure} 
\centering 
\includegraphics[width=87mm]{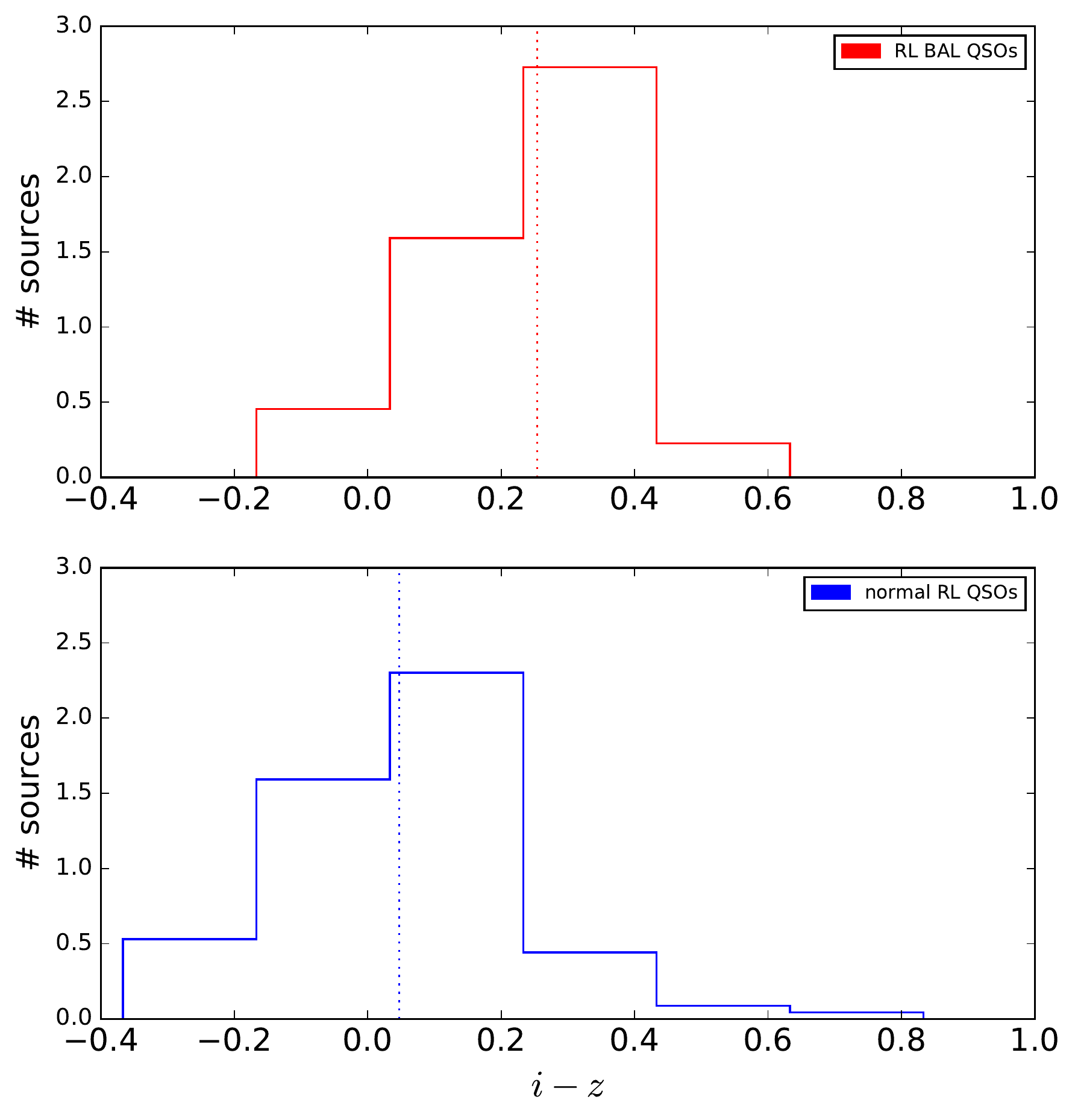} 
\caption{\small  Comparison of the normalized distribution of the (i-z) optical colors of high-z RL BAL and non-BAL QSOs, as discussed in sectio \ref{sec_optical}. The mean of the BAL QSOs colors is indicated by red a dashed lines, while for the normal RL QSOs is indicated by blue dashed lines.} 
\label{distributionColors3}
\end{figure}

 The same trend is analyzed in Fig. \ref{distributionColors2} and  \ref{distributionColors3}, where we plot the normalized distribution of the colors for which the t-test gives lower value for the statistic p, indicating a larger difference in the means of the colors for BAL and non-BAL QSOs.  

\subsection{Near- Mid- infrared colors} 
\label{sec_infrared} 

\begin{figure} 
\centering 
\includegraphics[width=85mm]{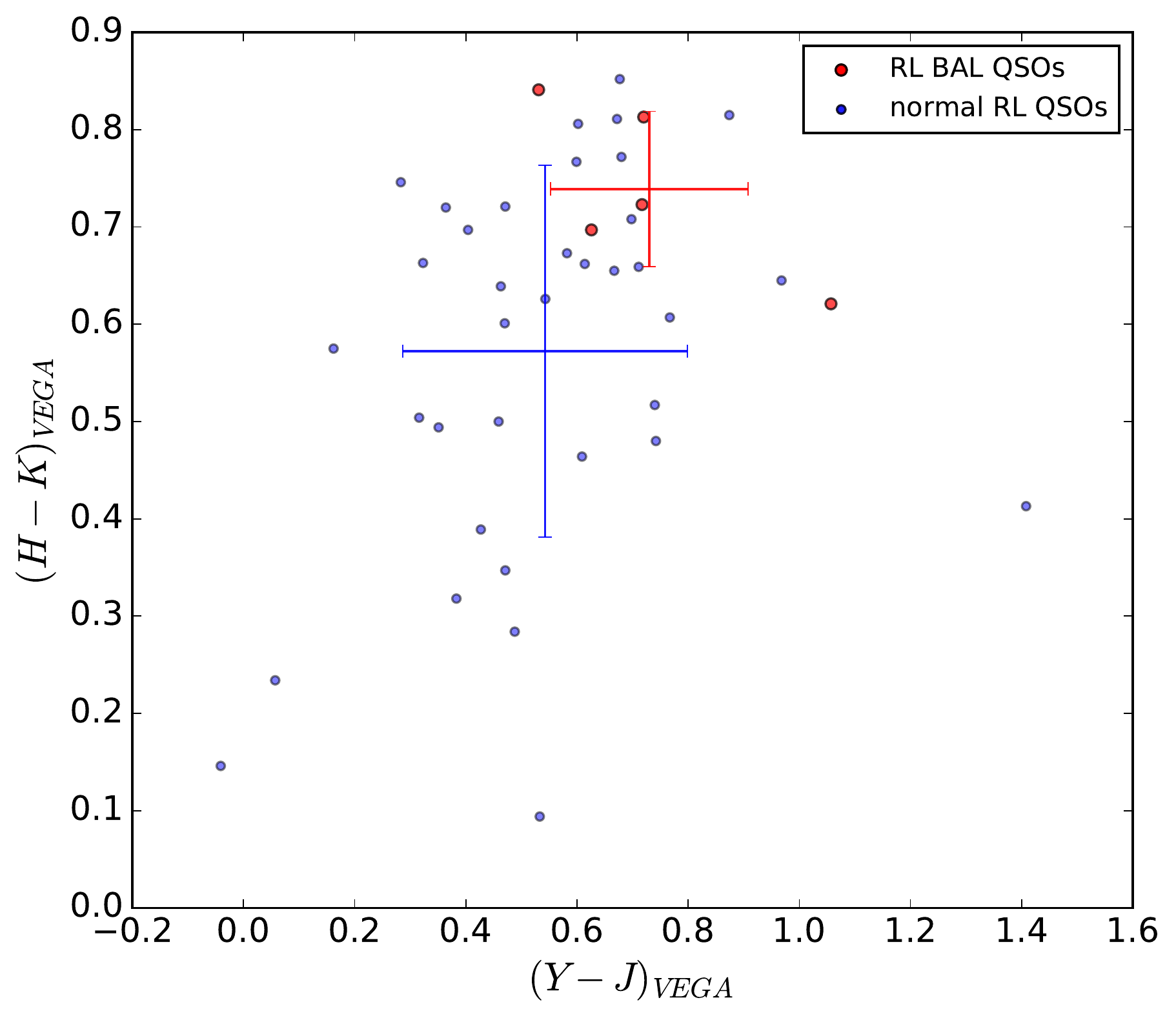} 
\caption{\small  Comparison of the NIR UKIDSS colors for a sample of 5 high-z RL BAL (red dots) and 36 normal RL QSOs (blue dots) matched in redshift. The means of the colors are shown in the plot as red cross for the BALs and as a blue cross for normal QSOs, both with 1-$\sigma$ error bar. See section \ref{sec_infrared}} 
\label{ukidss_colors}
\end{figure}

We matched our sample of 22  RL BAL QSOs and 113 normal QSOs with the LAS-DR9 UK Infrared Telescope Deep Sky Survey (UKIDSS; \citealt{Lawrence_2007}). 
For each source, we searched for the nearest counterpart within $2\farcs0$ from the radio-position (as given in FIRST). In fact, as shown \cite{Wu2010}, 99.6 per cent of the SDSS QSOs having matched counterpart in UKIDSS lie within 0.5 arcsec of the SDSS positions, and  the maximum radio-optical separation of the QSOs of our sample is less 1.0 arcsec . We have that 6 BAL QSOs and 49  normal QSOs lie in the footprint of UKIDSS DR9, however only 5 BAL QSOs and 36 normal QSOs are detected in all 4 UKIDSS bands (Y,J,H,K). In Fig. \ref{ukidss_colors} we show the comparison of the two samples in the $H-K$ vs $Y-J$ color-color space. The means of the UKIDSS colors are shown in the plot respectively as a red (for BAL QSOs) and a blue (for normal QSOs) cross. The t-test gives (see Table \ref{t_test2})  indicate that only the means of the latter colors are statistically different from each other.

\begin{figure} 
\centering 
\includegraphics[width=85mm]{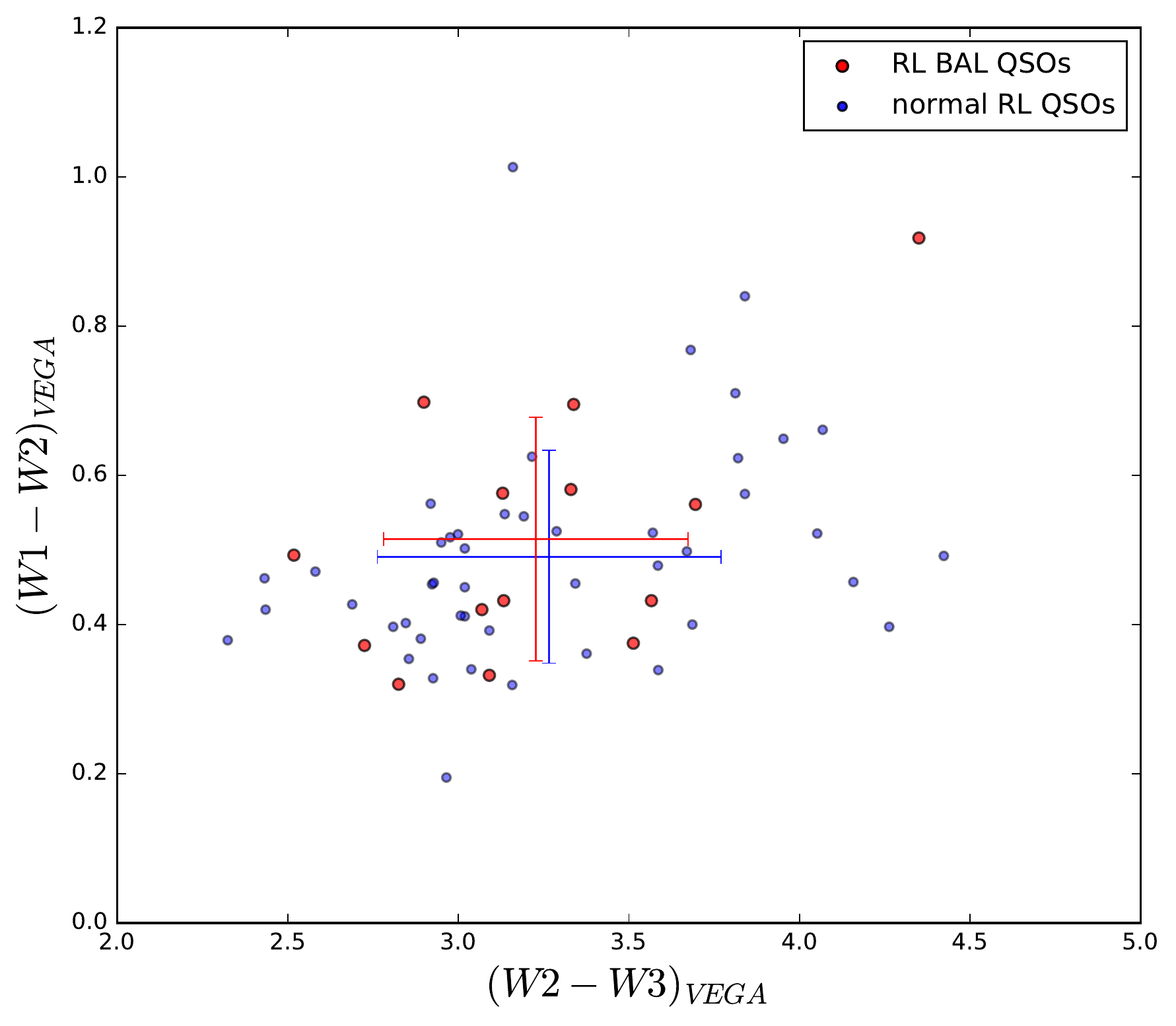} 
\caption{\small  Two-color diagram presenting the NIR WISE colors of 14 high-z RL BAL (red dots) and 47 normal RL QSOs (blue dots) matched in redshift. The large crosses with 1-$\sigma$ error bar, represent the location of the mean: red  for the BALs and as a blue for normal QSOs. See section \ref{sec_infrared}} 
\label{wise_colors}
\end{figure}

We also searched for counterparts of our sample of QSOs in the Wide-field Infrared Survey Explorer (WISE; \citealt{Wright_2010}) most recent data release (AllWISE; \citealt{Cutri_2013}) within  $\, 3\farcs0$ from the radio position. Thus, we used a slightly higher radius than the one of  $\, 2\farcs0$, used for the SDSS QSO Catalog (from the 9th DR, see \citealt{Paris_2012}) to cross match SDSS and WISE. In this case 104 normal QSOs and all 22 BAL QSOs are detected in all four (W1,W2,W3,W4) WISE bands. However we consider only the sources without image artifact contamination flags and  detected with a flux signal-to-noise ratio $>2$ in all W1,W2 and W3 bands. This way we consider two samples of 14 RL BAL QSOs and 47 normal RL BAL. Using the same methodology used for the SDSS and UKIDSS colors, we compare the WISE colors of the two sample as shown in Fig.\ref{wise_colors}. In this case, this simple comparison of the NIR colors does not reveal differences between the two samples. The means of the colors are compared with the t-test as resumed in Table \ref{t_test2} and none of these means are statistically different for the two samples. Thus we find no evidence of our high-z RL BAL QSOs having redder WISE colors than other optically selected RL QSOs at the same range of redshift.

\begin{table}
\centering
\label{t_test2}
\caption{t-test on NIR, MIR colors}
\renewcommand\tabcolsep{4pt}
\scalebox{0.90}{
\begin{tabular}{c c c c c c c c}
\hline \hline
\multicolumn{1}{c}{Color} &
\multicolumn{1}{c}{Population} &
\multicolumn{1}{c}{Mean} &
\multicolumn{1}{c}{$\sigma$} &
\multicolumn{1}{c}{Median} &
\multicolumn{1}{c}{t} &
\multicolumn{1}{c}{df} &
\multicolumn{1}{c}{p ($\times 10^{-2}$)} \\
\multicolumn{1}{c}{(1)} &
\multicolumn{1}{c}{(2)} &
\multicolumn{1}{c}{(3)} &
\multicolumn{1}{c}{(4)} &
\multicolumn{1}{c}{(5)} &
\multicolumn{1}{c}{(6)} &
\multicolumn{1}{c}{(7)} &
\multicolumn{1}{c}{(8)} \\
\hline
\multirow{2}{*}{(Y-J)}   & BAL         & 0.73  	& 0.18  & 0.72  & \multirow{2}{*}{1.90}  & \multirow{2}{*}{6.1} & \multirow{2}{*}{10.56}  \\
                                   & non BAL  & 0.54   & 0.26  & 0.54 &  \\
\hline
\multirow{2}{*}{(H-K)}   & BAL         &  0.74 	& 0.18	& 0.72 & \multirow{2}{*}{3.2}  & \multirow{2}{*}{10.44} & \multirow{2}{*}{0.83}  \\
                                   & non BAL  & 0.57	       & 0.26	& 0.63 &  \\
\hline
\multirow{2}{*}{(W2-W1)}   & BAL         & -0.51	& 0.16	& -0.46 & \multirow{2}{*}{0.5}  & \multirow{2}{*}{18.9} & \multirow{2}{*}{63.82}  \\
                                          & non BAL  & -0.49	& 0.14	& -0.46 &  \\
\hline
\multirow{2}{*}{(W3-W2)}   & BAL         & -3.23	& 0.45	& -3.13 & \multirow{2}{*}{0.3}  & \multirow{2}{*}{23.2} & \multirow{2}{*}{78.92}  \\
                                   	& non BAL  & -3.14	& 0.50	& -3.14 &  \\
\hline \hline                                                                     
\end{tabular}
}

\medskip
\begin{flushleft}
\small  \textit{The columns give the following: (1) Near and mid infrared color; (2) subsample considered; (3) mean of the color; (4) standard deviation; (5) median of the color; (6) statistic $t$; (7) associated degrees of freedom to the t-test; (8) the statistic p associated to the t-test, as usually, the null hypothesis of equal mean is rejected for values < 0.05}
\end{flushleft}
\end{table}

\section{Discussion}
In section \ref{radio_conclusions} we present the conclusions of our studies in the radio band, in section \ref{optical_conclusions} we present the conclusions of our investigations of optical and infrared colors of our RL BAL QSOs.

\subsection{Radio properties}
\label{radio_conclusions}

In the radio band, we studied a sample of 15 RL BAL QSOs and compared our results with the ones obtained for a well matched sample of 14 RL non-BAL QSOs. 

All objects are unresolved, even at the highest resolution observations performed with JVLA. The upper limit linear size deduced from the mean redshift of BAL and non-BAL samples is of $\sim$1.4 kpc, compatible with the one of radio sources in a early-phase (GPS and HFP). This is consistent with the synchrotron peak frequencies found, as discussed in the next paragraph.

We reconstructed the SED of our objects between 1.25 and 9.5 GHz. The fraction of young radio sources (i.e. GPS/HFP) in the BAL and non-BAL samples are compatible within errors, thus not suggesting a particular young radio phase for BAL QSOs with respect to \textit{normal} QSOs, even at this redshift range. A similar result was found at lower redshifts for the GPS fraction (\citealt{Bruni_2012}).

The fact that the orientation of BAL and non-BAL QSOs do not show a significant difference, is not in line with what found in lower redshift samples (\citealt{DiPompeo_2011}; \citealt{Bruni_2012}). This could be an effect of the higher fraction of GPS/HFP sources present in this work, due to the considered redshift window. In fact, if outflows responsible for the BAL absorption could be later re-oriented to form the jet (\citealt{Elvis}), young radio sources could present an outflows more likely oriented towards the polar direction, compensating for the tendency of BAL QSOs to have equatorial orientation. If verified with larger samples, this would link the jet-collimation mechanism with accretion-disk outflows reorientation, favoring the hypothesis that accretion disk rotation can be at the origin of magnetic jet launching (\citealt{Blandford}; \citealt{Boccardi}).


\subsection{Broadband colors}
\label{optical_conclusions}
 
 Numerous studies on the continuum and emission-line properties of BAL QSOs spectra have been pursued in the last 25 years, and it was soon noticed \citep{Weymann} that they show redder continua than those of normal quasars. This claim has been confirmed by a number of different studies (\citealt{Sprayberry_1992}; \citealt{Brotherton_2001}; \citealt{Reichard_2003}), and there is a general agreement on the fact that the subpopulation of LoBAL are significantly redder than HiBALs \citep{Urrutia_2009}, and that HiBAL are moderately redder than quasars not showing BAL features. However the origin of the reddening itself is still subject of debate (see for instance \citealt{Krawczyk_2015}) and the discussion is complicated from contrasting results in the near infrared (\citealt{Gallagher_2007}, \citealt{DiPompeo_2013}). In fact, if the differences in the mean optical-colors of BAL and non-BAL QSOs are consequence of a dustier environment (in agreement with the evolution scenario) they should be brighter in the infrared, where dust is seen in emission rather than in absorption.

In section \ref{secBrooadCol} we studied the SDSS optical color distributions of our sample of 22 RL BAL QSOs and of a sample of 113 normal RL QSOs with redshift in the same range and selected using the same criteria. The colors for BALQSOs and non-BALQSOs indicate that BALQSOs are redder than non-BALQSOs, and t-tests confirm that the means of the two groups are statistically different from each other. This result can not be due simply to absorption from the BAL troughs themselves, since the trough absorption can make the broadband color of BAL QSOs bluer as well as redder, depending on where the redshift of the quasar places the troughs with respect to the filters. Instead, it gives an indication of an overall flux deficit. These results confirm that BAL QSOs are redder than normal QSOs also at high redshift.

The UKIDSS colors of the subsample of 5 BAL and 36 non-BAL QSOs detected in this survey, indicate that the excess is likely to be extended at the  wavelength range of $0.83- 2.37 \, \mu m$  \citep{Hewett_2006}.  We extend the comparison at longest wavelengths considering the 14 BAL and the 47 non-BAL QSOs detected and with reliable photometry in WISE. However, in this case the comparison of the colors defined from the 3.4, 4.6 and 12$\mu m$ WISE bands, do not point to significant differences in the colors of BAL and non-BAL QSOs.

\section{Conclusions}
We have presented multi-frequency properties of the largest sample of RL BAL QSOs detected in SDSS DR7 and having $3.6 \le z \le 4.8$, i.e. the highest redshift bin that allow the identification of the BAL feature with optical spectra. The sample consist of 22 RL BAL QSOs,  4 of them identified as BAL QSOs in this work for the first time. We observed a fraction of them (15/22) in the radio band and we analyzed optical and infrared broadband colors of the whole sample. We can summarize the conclusions of this work as follows:

\begin{itemize}
\item All sources are unresolved, even when observed with the JVLA at 9 GHz (8 BAL \emph{vs} 8 non-BAL QSOs). This translates into
un upper limit for the projected linear size of 1.4 kpc, compatible with the GPS/HFP classification.
\item We compared the peak synchrotron frequencies for the BAL and non-BAL QSOs samples, not finding a predominance 
of GPS/HFP in the former. This does not suggest a particular younger radio phase for BAL QSOs with respect to non-BAL objects, 
even in this redshift range. Nevertheless, more than half of both samples can be classified as GPS/HFP, that is a larger higher fraction than the one found at lower redshift.  
\item We derived the spectral index for the two samples, and found that no statistically significant differences in orientation are present among BAL and non-BAL QSO objects. Given the high fraction of young radio sources present in this work (GPS/HFP), this could mean that BAL-producing outflows can have a preferential polar orientation in these objects, compensating the preferred equatorial orientation confirmed by different authors at lower redshifts. This would favor the hypothesis that  the jet can be collimated by accretion-disk driven magnetic force (\citealt{Blandford}; \citealt{Boccardi}), since outflow orientation and newly-formed jets would be connected. This should be verified on larger samples of GPS/HFP BAL QSOs.
\item We compare the broadband optical and NIR colors of our sample of 22 RL BAL QSOs and of 106 normal RL QSOs matched in redshift. We find that RL BALs QSOs BALQSOs tend to be located, on average, in redder regions of the color-color space respect to non-BAL RL QSOs. This trend is found in the optical (SDSS) and in the NIR wavelength of $0.83- 2.37 \, \mu m$ (UKIDSS). However we do not find significant differences in the two populations when comparing the colors at longer wavelength, i.e. at $3.4-12 \, \mu m$ (WISE).

\end{itemize}


\section*{Acknowledgements}
This work has been funded by the Spanish Ministerio de Economía y Competitividad (MINECO) under projects AYA2011-29517-C03-02 and AYA2014-58861-C3-2-P.
The research leading to these results has received funding from the European Commission Seventh Framework Programme (FP/2007-2013) under grant agreement No 283393 (RadioNet3).
This work is partially based on observations with the 100-m telescope of the MPIfR (Max-Planck-Institut f\"ur Radioastronomie) at Effelsberg. We thank the useful help from the Effelsberg operators. D. Tuccillo also thanks the University of Wyoming for hosting his useful and nice three months visit at their Department of Physics and Astronomy.
Finally, thank you to the anonymous referee, whose constructive comments assisted in clarifying and improving complex parts of the paper. 

The National Radio Astronomy Observatory is a facility of the National Science Foundation operated under cooperative agreement by Associated Universities, Inc. This research has made use of the NASA/IPAC Infrared Science Archive and NASA/IPAC Extragalactic Database (NED) which are both operated by the Jet Propulsion Laboratory, California Institute of Technology, under contract with the National Aeronautics and Space Administration. Use has been made of the Sloan Digital Sky Survey (SDSS) Archive. The SDSS is managed by the Astrophysical Research Consortium (ARC) for the participating institutions: The University of Chicago, Fermilab, the Institute for Advanced Study, the Japan Participation Group, The John Hopkins University, Los Alamos National Laboratory, the Max-Planck-Institute for Astronomy (MPIA), the Max-Planck-Institute for Astrophysics (MPA), New Mexico State University, University of Pittsburgh, Princeton University, the United States Naval Observatory, and the University of Washington.

\bibliography{mybib}

\begin{thebibliography}{62}
\expandafter\ifx\csname natexlab\endcsname\relax\def\natexlab#1{#1}\fi

\bibitem[{{Allen} {et~al}\mbox{.}(2011){Allen}, {Hewett}, {Maddox}, {Richards},
  \& {Belokurov}}]{Allen_2011}
{Allen} J.~T., {Hewett} P.~C., {Maddox} N., {Richards} G.~T., {Belokurov} V.,
  2011, \mnras, 410, 860

\bibitem[{{Antonucci}(1993)}]{Antonucci_1993}
{Antonucci} R., 1993, \araa, 31, 473

\bibitem[{{Baars} {et~al}\mbox{.}(1977){Baars}, {Genzel}, {Pauliny-Toth}, \&
  {Witzel}}]{Baars}
{Baars} J.~W.~M., {Genzel} R., {Pauliny-Toth} I.~I.~K., {Witzel} A., 1977,
  \aap, 61, 99

\bibitem[{{Balokovi{\'c}} {et~al}\mbox{.}(2012){Balokovi{\'c}}, {Smol{\v
  c}i{\'c}}, {Ivezi{\'c}}, {Zamorani}, {Schinnerer}, \&
  {Kelly}}]{Balokovic_2012}
{Balokovi{\'c}} M., {Smol{\v c}i{\'c}} V., {Ivezi{\'c}} {\v Z}., {Zamorani} G.,
  {Schinnerer} E., {Kelly} B.~C., 2012, ApJ, 759, 30

\bibitem[{{Becker} {et~al}\mbox{.}(1997){Becker}, {Gregg}, {Hook}, {McMahon},
  {White}, \& {Helfand}}]{Becker0}
{Becker} R.~H., {Gregg} M.~D., {Hook} I.~M., {McMahon} R.~G., {White} R.~L.,
  {Helfand} D.~J., 1997, \apj, 479, L93

\bibitem[{{Becker} {et~al}\mbox{.}(2000){Becker}, {White}, {Gregg},
  {Brotherton}, {Laurent-Muehleisen}, \& {Arav}}]{Becker1}
{Becker} R.~H., {White} R.~L., {Gregg} M.~D., {Brotherton} M.~S.,
  {Laurent-Muehleisen} S.~A., {Arav} N., 2000, \apj, 538, 72

\bibitem[{{Becker} {et~al}\mbox{.}(2001){Becker}, {White}, {Gregg},
  {Laurent-Muehleisen}, {Brotherton}, {Impey}, {Chaffee}, {Richards},
  {Helfand}, {Lacy}, {Courbin}, \& {Proctor}}]{Becker_2001}
{Becker} R.~H. {et~al.}, 2001, \apjs, 135, 227

\bibitem[{{Blandford} \& {Payne}(1982)}]{Blandford}
{Blandford} R.~D., {Payne} D.~G., 1982, \mnras, 199, 883

\bibitem[{{Boccardi} {et~al}\mbox{.}(2016){Boccardi}, {Krichbaum}, {Bach},
  {Bremer}, \& {Zensus}}]{Boccardi}
{Boccardi} B., {Krichbaum} T.~P., {Bach} U., {Bremer} M., {Zensus} J.~A., 2016,
  \aap, 588, L9

\bibitem[{{Boyle} {et~al}\mbox{.}(2000){Boyle}, {Shanks}, {Croom}, {Smith},
  {Miller}, {Loaring}, \& {Heymans}}]{Boyle_2000}
{Boyle} B.~J., {Shanks} T., {Croom} S.~M., {Smith} R.~J., {Miller} L.,
  {Loaring} N., {Heymans} C., 2000, MNRAS, 317, 1014

\bibitem[{{Brotherton} {et~al}\mbox{.}(2001){Brotherton}, {Tran}, {Becker},
  {Gregg}, {Laurent-Muehleisen}, \& {White}}]{Brotherton_2001}
{Brotherton} M.~S., {Tran} H.~D., {Becker} R.~H., {Gregg} M.~D.,
  {Laurent-Muehleisen} S.~A., {White} R.~L., 2001, \apj, 546, 775

\bibitem[{{Brotherton} {et~al}\mbox{.}(1998){Brotherton}, {van Breugel},
  {Smith}, {Boyle}, {Shanks}, {Croom}, {Miller}, \& {Becker}}]{Brotherton_1998}
{Brotherton} M.~S., {van Breugel} W., {Smith} R.~J., {Boyle} B.~J., {Shanks}
  T., {Croom} S.~M., {Miller} L., {Becker} R.~H., 1998, \apj, 505, L7

\bibitem[{{Bruni} {et~al}\mbox{.}(2013){Bruni}, {Dallacasa}, {Mack},
  {Montenegro-Montes}, {Gonz{\'a}lez-Serrano}, {Holt}, \&
  {Jim{\'e}nez-Luj{\'a}n}}]{Bruni_2013}
{Bruni} G., {Dallacasa} D., {Mack} K.-H., {Montenegro-Montes} F.~M.,
  {Gonz{\'a}lez-Serrano} J.~I., {Holt} J., {Jim{\'e}nez-Luj{\'a}n} F., 2013,
  \aap, 554, A94

\bibitem[{{Bruni} {et~al}\mbox{.}(2012){Bruni}, {Mack}, {Salerno},
  {Montenegro-Montes}, {Carballo}, {Benn}, {Gonz{\'a}lez-Serrano}, {Holt}, \&
  {Jim{\'e}nez-Luj{\'a}n}}]{Bruni_2012}
{Bruni} G. {et~al.}, 2012, \aap, 542, A13

\bibitem[{{Carballo} {et~al}\mbox{.}(2008){Carballo}, {Gonz{\'a}lez-Serrano},
  {Benn}, \& {Jim{\'e}nez-Luj{\'a}n}}]{Carballo_2008}
{Carballo} R., {Gonz{\'a}lez-Serrano} J.~I., {Benn} C.~R.,
  {Jim{\'e}nez-Luj{\'a}n} F., 2008, \mnras, 391, 369

\bibitem[{{Cutri et al.}(2013)}]{Cutri_2013}
{Cutri et al.}, 2013, VizieR Online Data Catalog, 2328

\bibitem[{{Dallacasa} {et~al}\mbox{.}(2000){Dallacasa}, {Stanghellini},
  {Centonza}, \& {Fanti}}]{Dallacasa}
{Dallacasa} D., {Stanghellini} C., {Centonza} M., {Fanti} R., 2000, \aap, 363,
  887

\bibitem[{{de Vries}, {Becker} \& {White}(2006){de Vries}, {Becker}, \&
  {White}}]{deVries_2006}
{de Vries} W.~H., {Becker} R.~H., {White} R.~L., 2006, \aj, 131, 666

\bibitem[{{DiPompeo}, {Brotherton} \& {De Breuck}(2012){DiPompeo},
  {Brotherton}, \& {De Breuck}}]{DiPompeo_2012}
{DiPompeo} M.~A., {Brotherton} M.~S., {De Breuck} C., 2012, \apj, 752, 6

\bibitem[{{DiPompeo} {et~al}\mbox{.}(2011){DiPompeo}, {Brotherton}, {De
  Breuck}, \& {Laurent-Muehleisen}}]{DiPompeo_2011}
{DiPompeo} M.~A., {Brotherton} M.~S., {De Breuck} C., {Laurent-Muehleisen} S.,
  2011, \apj, 743, 71

\bibitem[{{DiPompeo} {et~al}\mbox{.}(2013){DiPompeo}, {Runnoe}, {Brotherton},
  \& {Myers}}]{DiPompeo_2013}
{DiPompeo} M.~A., {Runnoe} J.~C., {Brotherton} M.~S., {Myers} A.~D., 2013,
  \apj, 762, 111

\bibitem[{{Elvis}(2000)}]{Elvis}
{Elvis} M., 2000, \apj, 545, 63

\bibitem[{{Fanti} {et~al}\mbox{.}(1990){Fanti}, {Fanti}, {Schilizzi},
  {Spencer}, {Nan Rendong}, {Parma}, {van Breugel}, \& {Venturi}}]{Fanti_1990}
{Fanti} R., {Fanti} C., {Schilizzi} R.~T., {Spencer} R.~E., {Nan Rendong},
  {Parma} P., {van Breugel} W.~J.~M., {Venturi} T., 1990, \aap, 231, 333

\bibitem[{{Fine}, {Jarvis} \& {Mauch}(2011){Fine}, {Jarvis}, \&
  {Mauch}}]{Fine_2011}
{Fine} S., {Jarvis} M.~J., {Mauch} T., 2011, \mnras, 412, 213

\bibitem[{{Gallagher} {et~al}\mbox{.}(2007){Gallagher}, {Hines}, {Blaylock},
  {Priddey}, {Brandt}, \& {Egami}}]{Gallagher_2007}
{Gallagher} S.~C., {Hines} D.~C., {Blaylock} M., {Priddey} R.~S., {Brandt}
  W.~N., {Egami} E.~E., 2007, \apj, 665, 157

\bibitem[{{Ganguly} {et~al}\mbox{.}(2007){Ganguly}, {Brotherton}, {Cales},
  {Scoggins}, {Shang}, \& {Vestergaard}}]{Ganguly_2007}
{Ganguly} R., {Brotherton} M.~S., {Cales} S., {Scoggins} B., {Shang} Z.,
  {Vestergaard} M., 2007, \apj, 665, 990

\bibitem[{{Gibson} {et~al}\mbox{.}(2009){Gibson}, {Jiang}, {Brandt}, {Hall},
  {Shen}, {Wu}, {Anderson}, {Schneider}, {Vanden Berk}, {Gallagher}, {Fan}, \&
  {York}}]{Gibson_2009}
{Gibson} R.~R. {et~al.}, 2009, \apj, 692, 758

\bibitem[{{Gregg} {et~al}\mbox{.}(1996){Gregg}, {Becker}, {White}, {Helfand},
  {McMahon}, \& {Hook}}]{Gregg_1996}
{Gregg} M.~D., {Becker} R.~H., {White} R.~L., {Helfand} D.~J., {McMahon} R.~G.,
  {Hook} I.~M., 1996, \aj, 112, 407

\bibitem[{{Hall} {et~al}\mbox{.}(2002){Hall}, {Anderson}, {Strauss}, {York},
  {Richards}, {Fan}, {Knapp}, {Schneider}, {Vanden Berk}, {Geballe}, {Bauer},
  {Becker}, {Davis}, {Rix}, {Nichol}, {Bahcall}, {Brinkmann}, {Brunner},
  {Connolly}, {Csabai}, {Doi}, {Fukugita}, {Gunn}, {Haiman}, {Harvanek},
  {Heckman}, {Hennessy}, {Inada}, {Ivezi{\'c}}, {Johnston}, {Kleinman},
  {Krolik}, {Krzesinski}, {Kunszt}, {Lamb}, {Long}, {Lupton}, {Miknaitis},
  {Munn}, {Narayanan}, {Neilsen}, {Newman}, {Nitta}, {Okamura}, {Pentericci},
  {Pier}, {Schlegel}, {Snedden}, {Szalay}, {Thakar}, {Tsvetanov}, {White}, \&
  {Zheng}}]{Hall_2002}
{Hall} P.~B. {et~al.}, 2002, \apjs, 141, 267

\bibitem[{{Hewett} {et~al}\mbox{.}(2006){Hewett}, {Warren}, {Leggett}, \&
  {Hodgkin}}]{Hewett_2006}
{Hewett} P.~C., {Warren} S.~J., {Leggett} S.~K., {Hodgkin} S.~T., 2006, \mnras,
  367, 454

\bibitem[{{Hopkins} \& {Elvis}(2010)}]{Hopkins_2010}
{Hopkins} P.~F., {Elvis} M., 2010, \mnras, 401, 7

\bibitem[{{Ivezi{\'c}} {et~al}\mbox{.}(2002){Ivezi{\'c}}, {Menou}, {Knapp},
  {Strauss}, {Lupton}, {Vanden Berk}, {Richards}, {Tremonti}, {Weinstein},
  {Anderson}, {Bahcall}, {Becker}, {Bernardi}, {Blanton}, {Eisenstein}, {Fan},
  {Finkbeiner}, {Finlator}, {Frieman}, {Gunn}, {Hall}, {Kim}, {Kinkhabwala},
  {Narayanan}, {Rockosi}, {Schlegel}, {Schneider}, {Strateva}, {SubbaRao},
  {Thakar}, {Voges}, {White}, {Yanny}, {Brinkmann}, {Doi}, {Fukugita},
  {Hennessy}, {Munn}, {Nichol}, \& {York}}]{Ivezic_2002}
{Ivezi{\'c}} {\v Z}. {et~al.}, 2002, \aj, 124, 2364

\bibitem[{{Jiang} {et~al}\mbox{.}(2007){Jiang}, {Fan}, {Ivezi{\'c}},
  {Richards}, {Schneider}, {Strauss}, \& {Kelly}}]{Jiang_2007}
{Jiang} L., {Fan} X., {Ivezi{\'c}} {\v Z}., {Richards} G.~T., {Schneider}
  D.~P., {Strauss} M.~A., {Kelly} B.~C., 2007, ApJ, 656, 680

\bibitem[{{Kellermann} {et~al}\mbox{.}(1989){Kellermann}, {Sramek}, {Schmidt},
  {Shaffer}, \& {Green}}]{Kellermann_1989}
{Kellermann} K.~I., {Sramek} R., {Schmidt} M., {Shaffer} D.~B., {Green} R.,
  1989, AJ, 98, 1195

\bibitem[{{Knigge} {et~al}\mbox{.}(2008){Knigge}, {Scaringi}, {Goad}, \&
  {Cottis}}]{Knigge_2008}
{Knigge} C., {Scaringi} S., {Goad} M.~R., {Cottis} C.~E., 2008, \mnras, 386,
  1426

\bibitem[{{Krawczyk} {et~al}\mbox{.}(2015){Krawczyk}, {Richards}, {Gallagher},
  {Leighly}, {Hewett}, {Ross}, \& {Hall}}]{Krawczyk_2015}
{Krawczyk} C.~M., {Richards} G.~T., {Gallagher} S.~C., {Leighly} K.~M.,
  {Hewett} P.~C., {Ross} N.~P., {Hall} P.~B., 2015, \aj, 149, 203

\bibitem[{{Lawrence} {et~al}\mbox{.}(2007){Lawrence}, {Warren}, {Almaini},
  {Edge}, {Hambly}, {Jameson}, {Lucas}, {Casali}, {Adamson}, {Dye}, {Emerson},
  {Foucaud}, {Hewett}, {Hirst}, {Hodgkin}, {Irwin}, {Lodieu}, {McMahon},
  {Simpson}, {Smail}, {Mortlock}, \& {Folger}}]{Lawrence_2007}
{Lawrence} A. {et~al.}, 2007, \mnras, 379, 1599

\bibitem[{{L{\'{\i}}pari} \& {Terlevich}(2006)}]{Lipari_2006}
{L{\'{\i}}pari} S.~L., {Terlevich} R.~J., 2006, \mnras, 368, 1001

\bibitem[{{Menou} {et~al}\mbox{.}(2001){Menou}, {Vanden Berk}, {Ivezi{\'c}},
  {Kim}, {Knapp}, {Richards}, {Strateva}, {Fan}, {Gunn}, {Hall}, {Heckman},
  {Krolik}, {Lupton}, {Schneider}, {York}, {Anderson}, {Bahcall}, {Brinkmann},
  {Brunner}, {Csabai}, {Fukugita}, {Hennessy}, {Kunszt}, {Lamb}, {Munn},
  {Nichol}, \& {Szokoly}}]{Menou_2001}
{Menou} K. {et~al.}, 2001, \apj, 561, 645

\bibitem[{{Montenegro-Montes} {et~al}\mbox{.}(2008){Montenegro-Montes}, {Mack},
  {Vigotti}, {Benn}, {Carballo}, {Gonz{\'a}lez-Serrano}, {Holt}, \&
  {Jim{\'e}nez-Luj{\'a}n}}]{Montenegro}
{Montenegro-Montes} F.~M., {Mack} K.-H., {Vigotti} M., {Benn} C.~R., {Carballo}
  R., {Gonz{\'a}lez-Serrano} J.~I., {Holt} J., {Jim{\'e}nez-Luj{\'a}n} F.,
  2008, \mnras, 388, 1853

\bibitem[{{Murray} {et~al}\mbox{.}(1995){Murray}, {Chiang}, {Grossman}, \&
  {Voit}}]{Murray_1995}
{Murray} N., {Chiang} J., {Grossman} S.~A., {Voit} G.~M., 1995, \apj, 451, 498

\bibitem[{{O'Dea}(1998)}]{ODea}
{O'Dea} C.~P., 1998, \pasp, 110, 493

\bibitem[{{Orr} \& {Browne}(1982)}]{Orr_1982}
{Orr} M.~J.~L., {Browne} I.~W.~A., 1982, \mnras, 200, 1067

\bibitem[{{P{\^a}ris} {et~al}\mbox{.}(2012){P{\^a}ris}, {Petitjean}, {Aubourg},
  {Bailey}, {Ross}, {Myers}, {Strauss}, {Anderson}, {Arnau}, {Bautista},
  {Bizyaev}, {Bolton}, {Bovy}, {Brandt}, {Brewington}, {Browstein}, {Busca},
  {Capellupo}, {Carithers}, {Croft}, {Dawson}, {Delubac}, {Ebelke},
  {Eisenstein}, {Engelke}, {Fan}, {Filiz Ak}, {Finley}, {Font-Ribera}, {Ge},
  {Gibson}, {Hall}, {Hamann}, {Hennawi}, {Ho}, {Hogg}, {Ivezi{\'c}}, {Jiang},
  {Kimball}, {Kirkby}, {Kirkpatrick}, {Lee}, {Le Goff}, {Lundgren}, {MacLeod},
  {Malanushenko}, {Malanushenko}, {Maraston}, {McGreer}, {McMahon},
  {Miralda-Escud{\'e}}, {Muna}, {Noterdaeme}, {Oravetz},
  {Palanque-Delabrouille}, {Pan}, {Perez-Fournon}, {Pieri}, {Richards},
  {Rollinde}, {Sheldon}, {Schlegel}, {Schneider}, {Slosar}, {Shelden}, {Shen},
  {Simmons}, {Snedden}, {Suzuki}, {Tinker}, {Viel}, {Weaver}, {Weinberg},
  {White}, {Wood-Vasey}, \& {Y{\`e}che}}]{Paris_2012}
{P{\^a}ris} I. {et~al.}, 2012, \aap, 548, A66

\bibitem[{{Proga}, {Stone} \& {Kallman}(2000){Proga}, {Stone}, \&
  {Kallman}}]{Proga_2000}
{Proga} D., {Stone} J.~M., {Kallman} T.~R., 2000, \apj, 543, 686

\bibitem[{{Reichard} {et~al}\mbox{.}(2003){Reichard}, {Richards}, {Hall},
  {Schneider}, {Vanden Berk}, {Fan}, {York}, {Knapp}, \&
  {Brinkmann}}]{Reichard_2003}
{Reichard} T.~A. {et~al.}, 2003, \aj, 126, 2594

\bibitem[{{Richards} {et~al}\mbox{.}(2011){Richards}, {Kruczek}, {Gallagher},
  {Hall}, {Hewett}, {Leighly}, {Deo}, {Kratzer}, \& {Shen}}]{Richards_2011}
{Richards} G.~T. {et~al.}, 2011, \aj, 141, 167

\bibitem[{{Rochais} {et~al}\mbox{.}(2014){Rochais}, {DiPompeo}, {Myers},
  {Brotherton}, {Runnoe}, \& {Hall}}]{Rochais_2014}
{Rochais} T.~B., {DiPompeo} M.~A., {Myers} A.~D., {Brotherton} M.~S., {Runnoe}
  J.~C., {Hall} S.~W., 2014, \mnras, 444, 2498

\bibitem[{{Saikia} \& {Salter}(1988)}]{Saikia_1988}
{Saikia} D.~J., {Salter} C.~J., 1988, \araa, 26, 93

\bibitem[{{Schlegel}, {Finkbeiner} \& {Davis}(1998){Schlegel}, {Finkbeiner}, \&
  {Davis}}]{Schlegel_1998}
{Schlegel} D.~J., {Finkbeiner} D.~P., {Davis} M., 1998, \apj, 500, 525

\bibitem[{{Schneider} {et~al}\mbox{.}(2010){Schneider}, {Richards}, {Hall},
  {Strauss}, {Anderson}, {Boroson}, {Ross}, {Shen}, {Brandt}, {Fan}, {Inada},
  {Jester}, {Knapp}, {Krawczyk}, {Thakar}, {Vanden Berk}, {Voges}, {Yanny},
  {York}, {Bahcall}, {Bizyaev}, {Blanton}, {Brewington}, {Brinkmann},
  {Eisenstein}, {Frieman}, {Fukugita}, {Gray}, {Gunn}, {Hibon}, {Ivezi{\'c}},
  {Kent}, {Kron}, {Lee}, {Lupton}, {Malanushenko}, {Malanushenko}, {Oravetz},
  {Pan}, {Pier}, {Price}, {Saxe}, {Schlegel}, {Simmons}, {Snedden}, {SubbaRao},
  {Szalay}, \& {Weinberg}}]{Schneider}
{Schneider} D.~P. {et~al.}, 2010, \aj, 139, 2360

\bibitem[{{Silk} \& {Rees}(1998)}]{Silk_1998}
{Silk} J., {Rees} M.~J., 1998, \aap, 331, L1

\bibitem[{{Sprayberry} \& {Foltz}(1992)}]{Sprayberry_1992}
{Sprayberry} D., {Foltz} C.~B., 1992, \apj, 390, 39

\bibitem[{{Stocke} {et~al}\mbox{.}(1992){Stocke}, {Morris}, {Weymann}, \&
  {Foltz}}]{Stocke_1992}
{Stocke} J.~T., {Morris} S.~L., {Weymann} R.~J., {Foltz} C.~B., 1992, \apj,
  396, 487

\bibitem[{{Tolea}, {Krolik} \& {Tsvetanov}(2002){Tolea}, {Krolik}, \&
  {Tsvetanov}}]{Tolea_2002}
{Tolea} A., {Krolik} J.~H., {Tsvetanov} Z., 2002, \apj, 578, L31

\bibitem[{{Trump} {et~al}\mbox{.}(2006){Trump}, {Hall}, {Reichard}, {Richards},
  {Schneider}, {Vanden Berk}, {Knapp}, {Anderson}, {Fan}, {Brinkman},
  {Kleinman}, \& {Nitta}}]{Trump}
{Trump} J.~R. {et~al.}, 2006, \apjs, 165, 1

\bibitem[{{Tuccillo}, {Gonz{\'a}lez-Serrano} \& {Benn}(2015){Tuccillo},
  {Gonz{\'a}lez-Serrano}, \& {Benn}}]{Tuccillo}
{Tuccillo} D., {Gonz{\'a}lez-Serrano} J.~I., {Benn} C.~R., 2015, \mnras, 449,
  2818

\bibitem[{{Urrutia} {et~al}\mbox{.}(2009){Urrutia}, {Becker}, {White},
  {Glikman}, {Lacy}, {Hodge}, \& {Gregg}}]{Urrutia_2009}
{Urrutia} T., {Becker} R.~H., {White} R.~L., {Glikman} E., {Lacy} M., {Hodge}
  J., {Gregg} M.~D., 2009, \apj, 698, 1095

\bibitem[{{Urry} \& {Padovani}(1995)}]{Urry_1995}
{Urry} C.~M., {Padovani} P., 1995, \pasp, 107, 803

\bibitem[{{Weymann} {et~al}\mbox{.}(1991){Weymann}, {Morris}, {Foltz}, \&
  {Hewett}}]{Weymann}
{Weymann} R.~J., {Morris} S.~L., {Foltz} C.~B., {Hewett} P.~C., 1991, \apj,
  373, 23

\bibitem[{{Wright} {et~al}\mbox{.}(2010){Wright}, {Eisenhardt}, {Mainzer},
  {Ressler}, {Cutri}, {Jarrett}, {Kirkpatrick}, {Padgett}, {McMillan},
  {Skrutskie}, {Stanford}, {Cohen}, {Walker}, {Mather}, {Leisawitz}, {Gautier},
  {McLean}, {Benford}, {Lonsdale}, {Blain}, {Mendez}, {Irace}, {Duval}, {Liu},
  {Royer}, {Heinrichsen}, {Howard}, {Shannon}, {Kendall}, {Walsh}, {Larsen},
  {Cardon}, {Schick}, {Schwalm}, {Abid}, {Fabinsky}, {Naes}, \&
  {Tsai}}]{Wright_2010}
{Wright} E.~L. {et~al.}, 2010, \aj, 140, 1868

\bibitem[{{Wu} \& {Jia}(2010)}]{Wu2010}
{Wu} X.-B., {Jia} Z., 2010, \mnras, 406, 1583

\end{thebibliography}
\bibliographystyle{mn2e}

\end{document}